\documentclass[11pt]{article}
\usepackage{jheppub}
 
\usepackage[compat=1.1.0]{tikz-feynman} 
\usepackage{subcaption} 
\usepackage{multirow} 
\usepackage{color} 
\usepackage{comment} 
\usepackage{tablefootnote} 

\graphicspath{
  {./}
  {./new-figures/}
}

\definecolor{mymagenta}{rgb}{1,0,1}

\newcommand\MSbar{$\overline{\rm MS}$}
\newcommand{\mc}[1]{\mathcal{#1}}

\newcommand{\mr}[1]{\mathrm{#1}}
\newcommand{\rmi}[1]{{\mbox{\scriptsize #1}}}
\newcommand{\rmii}[1]{{\mbox{\tiny\rm{#1}}}}

\newcommand{\rg}{\mu}
\newcommand{\drg}[1]{\rg \frac{d #1}{d\rg}}
\newcommand{\sumint}[1]{{\hbox{$\sum$}\!\!\!\!\!\!\!\int\,}_{\!\!\!\!\raise-0.9ex\hbox{$\scriptstyle{#1}$}}}

\newcommand{\gsq}{g^2}
\newcommand{\gsqsq}{g^4}
\newcommand{\gsqsqsq}{g^6}

\newcommand{\nn}{\nonumber}
\newcommand{\gammaE}{{\gamma}}

\newcommand{\Tint}[1]{{\hbox{$\sum$}\!\!\!\!\!\!\!\int\,}_{\!\!\!\!\raise-0.9ex\hbox{$\scriptstyle{#1}$}}}
\newcommand{\Tinti}[1]{{{\Sigma}\!\!\!\!\raise0.3ex\hbox{$\int$}_\rmii{${#1}$}}}
\newcommand{\Tintip}[1]{{{\Sigma'}\!\!\!\!\!\raise0.3ex\hbox{$\int$}_\rmii{${#1}$}}}

\newcommand{\treelevel}[1]{
    \vertex (a) at (0,0);
    \vertex (b) at (2,0);
    \diagram*{
    (a) -- [#1] (b)
    };
}
\newcommand{\selfenergyfour}[2]{
    \vertex (a) at (0,0);
    \vertex (b) at (1,0);
    \vertex (c) at (2,0);
    \diagram*{
    (a) -- [#1] (b) -- [#1] (c)
    };
    \draw[#2] (b) arc [start angle=-90, end angle=270, radius=0.7cm];
}
\newcommand{\sunset}[2]{
    \vertex (a) at (0,0);
    \vertex (b) at (1.5,0);
    \vertex (c) at (3,0);
    \vertex (d) at (1.5,-0.7);
    \diagram*{
    (a) -- [#1] (b) -- [#1] (c)
    };
    \draw[#2] (d) arc [start angle=-90, end angle=270, radius=0.7cm];
}
\newcommand{\cactus}[2]{
    \vertex (a) at (0,0);
    \vertex (b) at (1,0);
    \vertex (c) at (2,0);
    \vertex (d) at (1,1.4);
    \diagram*{
    (a) -- [#1] (b) -- [#1] (c)
    };
    \draw[#2] (b) arc [start angle=-90, end angle=270, radius=0.7cm];
    \draw[#2] (d) arc [start angle=-90, end angle=270, radius=0.7cm];
}

\title{On the perturbative expansion at high temperature and
implications for cosmological phase transitions}
\author[a]{Oliver Gould}
\author[b,c,d]{and Tuomas V.~I.~Tenkanen}

\preprint{NORDITA 2021-010}

\affiliation[a]{School of Physics and Astronomy, University of Nottingham, Nottingham NG7 2RD, United Kingdom}
\affiliation[b]{Nordita, KTH Royal Institute of Technology and Stockholm University, Roslagstullsbacken 23, SE-106 91 Stockholm, Sweden}
\affiliation[c]{Tsung-Dao Lee Institute \& School of Physics and Astronomy, Shanghai Jiao Tong University, Shanghai 200240, China}
\affiliation[d]{Shanghai Key Laboratory for Particle Physics and Cosmology, Key Laboratory for Particle Astrophysics and Cosmology (MOE), Shanghai Jiao Tong University, Shanghai 200240, China}

\emailAdd{oliver.gould@nottingham.ac.uk}
\emailAdd{tuomas.tenkanen@su.se}

\date{\today}
\abstract{
We revisit the perturbative expansion at high temperature and
investigate its convergence by inspecting the
renormalisation scale dependence of the effective potential.
Although at zero temperature the renormalisation group improved effective potential is scale independent at one-loop, we show how this breaks down at high temperature, due to the misalignment of loop and coupling expansions.
Following this, we show how one can recover renormalisation scale independence at high temperature,
and that it requires computations at two-loop order.
We demonstrate how this resolves some of the huge theoretical uncertainties in the gravitational wave signal of first-order phase transitions, though uncertainties remain stemming from the computation of the bubble nucleation rate.
}

\begin{document}

\maketitle

%

\section{Introduction}
\label{sec:intro}

The discovery of gravitational waves by LIGO~\cite{Abbott:2016blz} suggests a novel possibility to observe the early universe.
The comparatively weak interaction between gravity and matter means that gravitational waves will free-stream after their production, carrying with them a fingerprint of whatever process produced them, a fingerprint which is potentially observable today.
In particular, a strong first-order phase transition in the early universe would give rise to a stochastic background of gravitational waves (for recent reviews see Refs.~\cite{Caprini:2018mtu,Caprini:2019egz,Hindmarsh:2020hop}), which may be visible by planned gravitational wave experiments such as LISA~\cite{Audley:2017drz}, DECIGO~\cite{Kawamura:2011zz}, BBO~\cite{Harry:2006fi} and Taiji~\cite{Guo:2018npi}.
In fact recent evidence for a possible stochastic background of gravitational waves by the NANOGrav experiment~\cite{Arzoumanian:2020vkk} has been interpreted as the gravitational wave signal of a first-order phase transition~\cite{Addazi:2020zcj,Nakai:2020oit,Li:2020cjj,Ratzinger:2020koh}.

On the other hand, the discovery of the Higgs boson~\cite{Chatrchyan:2012ufa,Aad:2012tfa} and high precision probes of its interactions at the LHC and planned future high-energy colliders play a central role in advancing our understanding of particle physics.
In particular, determining the symmetry-breaking pattern of the electroweak sector is a major goal.
A first-order electroweak phase transition may provide the necessary departure from equilibrium for the generation of the observed matter-antimatter asymmetry \cite{Morrissey:2012db}, yet this possibility requires new particles with masses $\lesssim \mathrm{TeV}$ which are not too weakly coupled to the Higgs \cite{Ramsey-Musolf:2019lsf}.
The determination of the pattern of electroweak symmetry breaking is therefore a clear target for near-future collider experiments.

From an observation of a stochastic background of gravitational waves, one can in principle learn about the underlying physics of the production process.
Indeed, it has been mooted that this may provide an experimental probe of particle physics beyond the Standard Model (BSM) which may be complementary and competitive with collider searches,
see for example Refs.~\cite{Kozaczuk:2019pet,Ramsey-Musolf:2019lsf} and references therein.
However, for this to be feasible, it must be possible to make robust quantitative predictions of the gravitational wave spectrum produced by a given particle physics model.
For first-order phase transitions, this in turn requires accurate predictions of the thermodynamics of the phase transition.

The majority of recent literature studying the thermodynamics of cosmological first-order phase transitions -- in a wide variety of models -- resorts to \textit{one-loop} computations that use the (daisy-resummed) thermal effective potential.
Such studies are aided by the known explicit form of the potential (in terms of background field-dependent mass eigenvalues) in this approximation
\cite{Quiros:1999jp,Cline:2006ts,Patel:2011th,Wainwright:2011kj,Morrissey:2012db,Basler:2018cwe,Athron:2020sbe}, 
so allowing for relatively straightforward applications even to complicated BSM models.
This simplicity has led to a wealth of phenomenological studies, for example 
\cite{Huber:2000mg,Grojean:2004xa,Ham:2004cf,Bodeker:2004ws,Fromme:2006cm,Delaunay:2007wb,Espinosa:2007qk,Profumo:2007wc,Noble:2007kk,Espinosa:2008kw,Funakubo:2009eg,Cline:2009sn,Kehayias:2009tn,Espinosa:2010hh,Espinosa:2011ax,Gil:2012ya,Chung:2012vg,Leitao:2012tx,Dorsch:2013wja,Profumo:2014opa,Curtin:2014jma,Jiang:2015cwa,Blinov:2015sna,Kozaczuk:2015owa,Vaskonen:2016yiu,Basler:2016obg,Beniwal:2017eik,Chiang:2017nmu,Basler:2017uxn,Chala:2018ari,Dev:2019njv}
which have broadened our understanding of the possible scope of planned gravitational wave experiments to probe particle physics (as well as viability of electroweak baryogenesis).
However, as recently emphasised in Refs.~\cite{Croon:2020cgk,Papaefstathiou:2020iag}, such one-loop computations suffer from huge theoretical uncertainties due to thermal enhancements of infrared physics, significantly limiting the possibility of making quantitative conclusions.
These amount to orders of magnitude uncertainty in the peak amplitude of the gravitational wave spectrum.%
\footnote{
In fact, the theoretical uncertainties may be larger still than those found in Ref.~\cite{Croon:2020cgk}.
For example, Ref.~\cite{Carena:2019une} compared two different one-loop approximations within the $Z_2$-symmetric xSM, and observed a difference of ten orders of magnitude.
}

Unphysical dependence on the renormalisation scale ($\mu$) was noted as one of the largest theoretical uncertainties in typical one-loop computations~\cite{Croon:2020cgk}.
This is in stark contrast with typical one-loop computations at zero temperature, in which renormalisation scale dependence is usually small, at least for theories with small couplings.
As a consequence, this issue has been the subject of confusion in the literature. Incomplete attempts at renormalisation group (RG) improvement of the thermal effective potential were made for example in
Refs.~\cite{Espinosa:1995se,Rose:2015lna,Blinov:2015sna,Huber:2015znp,Cai:2017tmh,Chiang:2017nmu,Braconi:2018gxo,Chiang:2019oms,Carena:2021onl}.
Note that renormalisation scale dependence can be regarded as a proxy revealing the size of missing higher-order terms; a strong renormalisation scale dependence reveals important missing terms.

\begin{figure}[t]
    \centering
    \includegraphics[width=.65\textwidth]{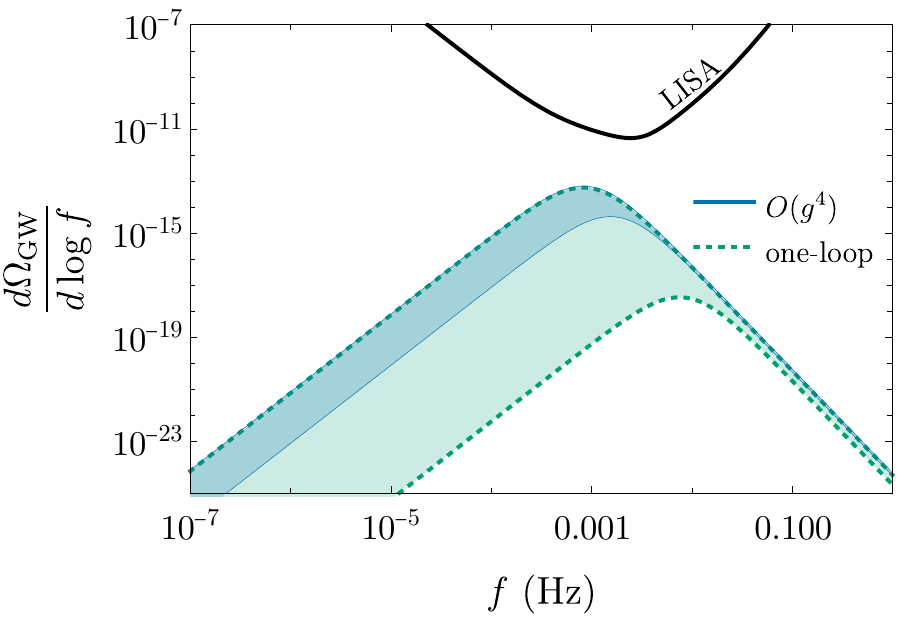}
    \caption{
    The gravitational wave spectrum produced by a first-order phase transition in the $Z_2$-symmetric xSM, together with LISA Science Requirements sensitivity curve~\cite{LISAScienceRequirements} in black.
    The coloured lines and associated theoretical uncertainty bands show the predicted gravitational wave spectrum for a single benchmark parameter point in this model (labelled BM1 in Sec.~\ref{sec:equilibrium}).
    The uncertainty bands reflect the variation as a result of varying the renormalisation scale over a factor of 4 for these two different approximations.
    For more details see Sec.~\ref{sec:gw}.
    The LISA signal-to-noise ratio, assuming a 3 year mission profile and using {\tt PTPlot}~\cite{Caprini:2019egz} for the computation, varies from 0.017 to 0.19 in the $\mc{O}(g^4)$ approximation, and from $2.3\times 10^{-6}$ to 0.17 in the one-loop approximation.
    }
    \label{fig:gws-BM1}
\end{figure}

In Fig.~\ref{fig:gws-BM1} we show an example of the renormalisation scale dependence of the gravitational wave spectrum from a first-order phase transition, in the real singlet-extended Standard Model (xSM).
Shown are predictions for one benchmark parameter point based on two different approximations:
the widely used one-loop thermal effective potential in (dotted) green,
and a more complete computation in blue.
In the one-loop calculation, both the peak amplitude of the gravitational wave spectrum and the LISA signal-to-noise ratio vary by four orders of magnitude as the renormalisation scale varies over just a factor of four, a huge theoretical uncertainty.
This intrinsic uncertainty is much smaller for the more complete computation%
\footnote{
Nevertheless, it is still an order of magnitude, demonstrating the importance of additional higher-order contributions.
}
that is organised in powers of couplings (rather than the loop expansion) and accounts for (equilibrium) contributions consistently up to $\mc{O}(g^4)$.
Here $g$ is a formal power counting parameter, that is identified with the SU(2) gauge coupling in the case of the xSM.
For a detailed discussion of this analysis, see Sec.~\ref{sec:gw}.       

As we show in this article, for $Z_2$-symmetric theories, the trouble arises from the implicit running of thermal corrections to quadratic terms in the potential%
\footnote{For non-$Z_2$-symmetric theories there is a similar troublesome thermal contribution to the tadpole.}
\begin{equation}
\frac{1}{2} \Big( m^2 + \Pi_T \Big) v^2, \label{eq:troublesome-term}
\end{equation}
leading to the strong renormalisation scale dependence of the high temperature effective potential.
Here $v$ is a real background field, $m^2$ is the zero temperature mass parameter and $\Pi_T$ is the one-loop thermal contribution to the self-energy, i.e.\ the one-loop thermal mass. 
Denoting the interaction coupling by $g^2$, the thermal contribution to the self-energy has the schematic form $\Pi_T \sim g^2 T^2$.
The implicit running of the $\Pi_T v^2$ term is an effect at $\mathcal{O}(g^4 T^2 v^2)$ and is not cancelled by any one-loop, or resummed one-loop, contribution to the effective potential.
It is cancelled by explicit logarithms which only appear at \textit{two-loop} level.

Thus the cancellation of renormalisation scale dependence requires more work at high temperature than it does at zero temperature.
This is a specific case of a more general feature of physics at high temperatures: due to the high occupation of infrared modes, the effective expansion parameter is the square root of the coupling, necessitating higher-order calculations.
While the one-loop approximation may suffice at zero temperature, one must work harder at high temperatures to achieve the same degree of accuracy.

The realisation that thermal effects can correct the effective potential at leading $\mc{O}(g^2)$ order goes back to Ref.~\cite{Dolan:1973qd}, where the thermal correction in Eq.~\eqref{eq:troublesome-term} was derived.
The more general misalignment of the loop expansion and the coupling expansion at high temperature was further clarified in Ref.~\cite{Kapusta:1979fh}, in which the presence of a contribution at $\mc{O}(g^3)$ was demonstrated.
However, the running of the tree-level potential starts at $\mc{O}(g^4)$, so it was not until calculations were pushed to this order~\cite{Parwani:1991gq,Arnold:1992rz} that issues of renormalisation scale dependence were directly tackled.

Ref.~\cite{Arnold:1992rz} pioneered the perturbative study of electroweak-type theories at high temperature.
The approach was based upon organising perturbation theory as an expansion in powers of couplings, and not by the loop expansion.
Strictly adopting such an expansion, the authors established that there is a unique way to handle the infamous ring or daisy diagrams, thereby fixing the $\mc{O}(g^3)$ term in the effective potential.
Furthermore, the computation was extended to $\mc{O}(g^4)$.
This required a two-loop computation, and the authors demonstrated that their results are -- correctly -- renormalisation group invariant to this order.
In Refs.~\cite{Fodor:1994bs,Buchmuller:1995sf} this calculation was extended, allowing for a parametrically larger Higgs self-coupling.

A more detailed discussion of renormalisation scale dependence at high temperature was given in Ref.~\cite{Farakos:1994kx}, formulated in the framework of {\em (high temperature) dimensional reduction}, a framework that was at least implicit in Ref.~\cite{Arnold:1992rz}.
In this, an important distinction was made between the renormalisation scale dependence associated with the heavy non-zero Matsubara modes, and that associated with the light zero Matsubara modes.
Dimensional reduction was further developed in Refs.~\cite{Kajantie:1995dw,Braaten:1995cm,Farakos:1994xh} (for reviews see Refs.~\cite{Andersen:2004fp,Laine:2016hma,Ghiglieri:2020dpq,Schicho:2021gca}).
An early triumph of the approach was the determination of the phase diagram of the Standard Model electroweak sector~\cite{Kajantie:1995kf,Kajantie:1996mn,Kajantie:1996qd}.
Building on these earlier works, later studies using dimensional reduction to study hot electroweak theories have typically performed complete $\mc{O}(g^4)$ computations, for which the cancellation of renormalisation scale dependence serves as a useful crosscheck 
(c.f.\ for example~\cite{Andersen:1998br,Losada:1998at,Losada:1999tf,Laine:2000kv,Gynther:2003za,Gynther:2005dj,Gynther:2005av,Laine:2015kra,Gorda:2018hvi,Niemi:2018asa, Kainulainen:2019kyp,Niemi:2020hto,Gould:2021dzl,Niemi:2021qvp}).

The majority of recent literature on cosmological first-order phase transitions utilises the daisy resummation scheme developed in Ref.~\cite{Arnold:1992rz}, but falls short of carrying out a complete calculation at $\mc{O}(g^4)$, in particular missing logarithms of the renormalisation scale from two-loop thermal masses and vacuum diagrams (as well as logarithms related to field renormalisation).
Hence, the results of these studies exhibit a strong renormalisation scale dependence, which cannot be removed by RG running of tree-level and one-loop terms, due to the incompleteness of the computation at $\mc{O}(g^4)$. 

In this article, we aim to clarify 
the structure of the perturbative expansion at high temperature,
as revealed through the lens of renormalisation scale dependence.
We focus on the effective potential, as this carries much of the important thermodynamic information, though our general conclusions are not limited to the effective potential.
The remainder of this article is organised as follows.
To illustrate the crucial issues without unnecessary complications, we start by working with a simple example: the $Z_2$-symmetric real scalar field, or $\phi^4$-theory.
In Sec.~\ref{sec:rg-zero-T} we review the RG improvement and scale independence of the effective potential at zero temperature.
In Sec.~\ref{sec:rg-high-T} we show how this fails at high temperatures, leaving an uncancelled renormalisation scale dependence for the one-loop thermal effective potential.
Further in Sec.~\ref{sec:rg-impr-high-T} we show how this can be resolved by adding the leading two-loop contributions, which are $\mathcal{O}(g^4)$ in the coupling expansion. 
In Sec.~\ref{sec:equilibrium} we turn to a realistic candidate theory of particle physics, the xSM, and demonstrate the numerical importance of our conclusions for predictions of
equilibrium thermodynamics. 
In Sec.~\ref{sec:gw} we extend this discussion to the gravitational wave spectrum.
Finally, we summarise and conclude in Sec.~\ref{sec:discussion}.
For completeness, Appendix~\ref{sec:details} collects some explicit technical details of the computation.

\section{Scale independence at zero temperature}
\label{sec:rg-zero-T}

To demonstrate issues of scale dependence without unnecessary complications, we initially work with the simplest possible theory,
that of a single real scalar field, $\phi$, with a $Z_2$ symmetry, $\phi\to -\phi$.
The Lagrangian density is
\begin{align}
\mc{L} &= \frac{1}{2}\partial_\mu \phi \partial^\mu \phi -V(\phi), \label{eq:lagrangian} \\
V(\phi) 
&= \frac{1}{2}m^2(\rg)\phi^2 + \frac{1}{4!}\gsq(\rg) \phi^4. \label{eq:potential} 
\end{align}
We regularise the theory using dimensional regularisation, and choose our counterterms 
according to the \MSbar\ scheme, see Appendix \ref{sec:details} .
The arbitrary renormalisation scale introduced in this way is denoted by $\rg$, and the parameters $m^2(\rg)$ and $\gsq(\rg)$ are the corresponding renormalised \MSbar\ parameters. 
We denote the scalar self-coupling by $g^2$, in analogy with gauge theories.
In the following, we will not always show the explicit argument $\rg$, but one should keep this renormalisation scale dependence in mind.
We will assume throughout that the theory is perturbative, so that the loop-expansion parameter is small, $\gsq/(4\pi)\ll 1$.

At zero temperature, the phase structure of the theory can be determined from the minima of the perturbatively computed effective potential.
To calculate this, one shifts the field by a constant, homogeneous background $\phi \to v + \phi$, and then carries out a loop expansion for the fluctuating field $\phi$, dropping linear terms~\cite{Jackiw:1974cv}.
At leading order, the result is just the tree-level potential evaluated on the background field,
\begin{equation}
\label{eq:Vtree}
V_\rmi{tree} (v) = \frac{1}{2}m^2v^2 + \frac{1}{4!}\gsq v^4.
\end{equation}
For positive $m^2>0$ this has a minimum at $v=0$, whereas for negative $m^2<0$ this has two minima at $v^2=-m^2/\gsq$.

The perturbatively computed potential depends on the renormalisation scale, $\rg$, through the Lagrangian \MSbar\ parameters.
For a generic background field, the leading order scale dependence takes the form%
\footnote{Due to the absence of momentum-dependent divergences at one-loop order in this theory, there is no anomalous dimension; see Eq.~\eqref{eq:rge_appendix}.}
\begin{align}
\drg{} V_\rmi{tree} (v) &= \frac{1}{2}\drg{m^2}v^2 + \frac{1}{4!}\drg{\gsq} v^4, \\
&= \frac{\gsq}{(4\pi)^2}\left(\frac{1}{2}m^2v^2 + \frac{1}{8} \gsq v^4\right), \label{eq:tree-scale-dependence}
\end{align}
where we have used the one-loop beta functions, collected in Eqs.~\eqref{eq:betas}.

At one-loop, the effective potential is corrected by fluctuations of the $\phi$ field, yielding the Coleman-Weinberg contribution~\cite{Coleman:1973jx} (see also Appendix~\ref{sec:details}),
\begin{align}
\label{eq:VCWbar}
V_\rmi{CW}(v) &= \frac{1}{4(4\pi)^2}M^4(v)
\left[
\log \left(\frac{ M^2(v)}{\rg^2}\right)-\frac{3}{2}
\right] + \mr{const},\\
M^2(v) &= m^2 +\frac{1}{2} \gsq v^2. \label{eq:M_definition}
\end{align}
Note the presence of the renormalisation scale $\rg$ explicitly within the potential.
Here we use the freedom to shift the potential energy by a constant to fix $V(0)=0$ for the full one-loop potential.

The explicit scale dependence of the one-loop piece of the potential is
\begin{align}
\drg{} V_\text{CW} (v) = \frac{\gsq}{(4\pi)^2}\left(-\frac{1}{2}m^2v^2 - \frac{1}{8} \gsq v^4\right).
\end{align}
As one can see, this cancels the scale dependence of the parameters in the tree-level potential, shown in Eq.~\eqref{eq:tree-scale-dependence},
\begin{equation}
\drg{} \left(V_\rmi{tree}(v) + V_\rmi{CW} (v) \right) = 0. \label{eq:zero-scale-dependence}
\end{equation}
The cancellation however only holds at leading order, receiving corrections from the one-loop running of parameters within the one-loop potential, and from two-loop corrections to the running of parameters in the tree-level potential. However these are suppressed relatively by $\gsq/(4\pi)$,
and hence can be neglected in a one-loop analysis.

In summary, at zero temperature the effective potential is independent of the renormalisation scale at one-loop order.
In fact this holds order-by-order in $\hbar$,%
\footnote{
The reduced Planck's constant, $\hbar$, appears as the loop counting parameter, though it is set to unity in natural units.
}
or equivalently in $\gsq$ for this theory.
Thus renormalisation scale dependence of the zero temperature effective potential is always a higher order effect that can be neglected.
The inclusion of the running of couplings within the potential is often called renormalisation group (RG) improvement, see for example Refs.~\cite{Kastening:1991gv,Bando:1992np,Ford:1992mv}.
Our analysis has been carried out in the \MSbar\ renormalisation scheme, but the conclusions hold independently of this because the one-loop (and two-loop) beta functions, as well as the logarithm of the Coleman-Weinberg potential, are independent of the renormalisation scheme~\cite{Peskin:1995ev}.

\section{Scale dependence at high temperature}
\label{sec:rg-high-T}

At high temperature, the infrared modes of bosons become highly occupied.
As a consequence their effective coupling becomes larger than at zero temperature.
This causes a misalignment of the loop expansion and the expansion in powers of couplings, necessitating resummation of infinite classes of diagrams.
For a recent review, see Chapters~3 and 6 of Ref~\cite{Laine:2016hma}.

In this section, we will investigate the renormalisation scale dependence of the resummed one-loop effective potential.
As we will show, due to the enhancement of infrared modes, the cancellation of scale dependence which occurs at zero temperature 
no longer occurs at high temperature.
Thus, at high-temperature, the resummed one-loop effective potential is strongly scale dependent.

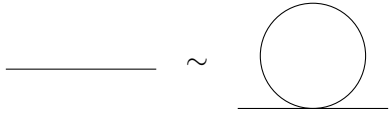
\begin{figure}[t]
    \centering
\begin{align*}
\begin{gathered}
	\begin{tikzpicture}[scale=1, transform shape]
    \begin{feynman}
    \treelevel{plain}
    \end{feynman}
    \end{tikzpicture}
\end{gathered}
\quad {\huge{\sim}} \quad
\begin{gathered}
	\begin{tikzpicture}[scale=1, transform shape]
    \begin{feynman}
    \selfenergyfour{plain}{plain}
    \end{feynman}
    \end{tikzpicture}
\end{gathered}
\end{align*}
    \caption{\label{fig:LO} At high temperature, the one-loop correction to the self-energy, or thermal mass, is of the same order 
    as the tree-level mass.
    Both terms are of order $\mc{O}(g^2T^2)$. 
    We use the {\tt TikZ-Feynman} package \cite{Ellis:2016jkw} to draw the diagrams.
    }
    \label{fig:one-loop-tree}
\end{figure}

At high-temperature thermal fluctuations become important, so that the tree-level potential is no longer the leading order approximation to the effective potential.
As has long been known, near the critical temperature of a phase transition, the dominant thermal corrections are to the mass, and these compete with the tree-level mass~\cite{Dolan:1973qd}.
Here and throughout we adopt the high-temperature approximation for thermal functions.
For the $Z_2$-symmetric scalar theory, the leading thermal contribution to the self-energy takes the form
\begin{equation}
\Pi_T = \frac{\gsq T^2}{24}.
\end{equation}
Near a phase transition it must be that $m^2$ and $\Pi_T$ are the same order of magnitude, so that
\begin{equation}
m^2 \sim \gsq T^2. \label{eq:power-counting}
\end{equation}
This is illustrated in Fig~\ref{fig:LO}.
At such high temperatures, the loop expansion is no longer appropriate.
Instead, the effective potential is computed as a power series in $\gsq$.
For strong symmetry-breaking transitions, strong enough to provide the necessary departure from equilibrium for successful electroweak baryogenesis, there is the further parametric relation $v \sim T$.
For ease of counting powers, we will often use this parametric relation when discussing the magnitude of different terms in the effective potential.
However, we will not assume this relation in our analysis.%
\footnote{
The phase transition in this $Z_2$-symmetric model is of second order~\cite{ZinnJustin:2002ru,Sun:2002cc}.
However, we will mostly refer to first-order phase transitions as these are of more interest to us.
Our general conclusions regarding renormalisation scale dependence however will not depend on the details of the transition, and as we show later in this section, neither will they depend on the model.
}

The thermal effective potential at leading order in $\gsq$ is instead,
\begin{equation}
V_{\text{thermal}}^{g^2} (v) = \frac{1}{2}\left(m^2+\Pi_T\right)v^2 + \frac{1}{4!}\gsq v^4.
\end{equation}
It is thus of order $\mc{O}(\gsq T^4)$ for $v\sim T$.
Note that $v\sim T$ also follows from demanding that the different terms of the effective potential are of the same order, $\Pi_T v^2 \sim \gsq v^4$, as one would expect in the vicinity of a phase transition.

In contrast to the effective potential, the renormalisation group equations are unaffected by finite temperature~\cite{Kapusta:2006pm}, being only dependent on the deep ultraviolet (UV).
However, because the effective potential is modified at leading order, so too is its running,
\begin{align}
\drg{} V_{\text{thermal}}^{g^2} (v) &= \frac{1}{2}\drg{m^2}v^2 + \frac{1}{2}\drg{\Pi_T}v^2 + \frac{1}{4!}\drg{\gsq} v^4, \\
&= \frac{1}{(4\pi)^2}\left(
\frac{1}{2} \gsq  m^2 v^2
+ \frac{1}{16} \gsqsq T^2 v^2
+ \frac{1}{8} \gsqsq v^4\right). \label{eq:lo-scale-dependence}
\end{align}

At subleading orders, half-integer powers of the coupling constant arise~\cite{Kapusta:1979fh}. This feature is a further consequence of the high occupation of infrared modes, unique to perturbation theory at high temperature.
The one-loop contribution of the zero Matsubara mode, appropriately daisy-resummed~\cite{Arnold:1992rz}, gives the first correction to the leading thermal effective potential, suppressed relatively by $\sqrt{\gsq}$.
This amounts to,
\begin{align}
V_{\text{thermal}}^{g^3} (v) = -\frac{T}{12\pi}\left(M^2(v)+\Pi_T\right)^{3/2} +\ \mr{const}, \label{eq:daisy}
\end{align}
where again we add a (temperature-dependent) constant to fix $V(0)=0$.
We do this because we are interested in differences between the two phases, and not in the absolute value of the free energy or pressure.
Eq.~\eqref{eq:daisy} is a correction to the effective potential of order 
$\mc{O}(g^3 T^4)$
for $v\sim T$.

Perhaps the most common approximation taken in the literature is a resummed one-loop approximation, in which thermal corrections to the effective potential are incorporated to one-loop order together with a resummation of daisy diagrams~\cite{Delaunay:2007wb,Espinosa:2007qk,Profumo:2007wc,Noble:2007kk,Espinosa:2008kw,Espinosa:2011ax,Curtin:2014jma,Blinov:2015sna,Basler:2016obg,Basler:2017uxn,Chala:2018ari}.%
\footnote{
Note that all these references use Arnold-Espinosa type daisy-resummation \cite{Arnold:1992rz}, where only the mass of the zero Matsubara mode is resummed, contrary to the approach by Parwani~\cite{Parwani:1991gq} in which all modes are resummed.
}
For example, this approach has been adopted in numerical packages for analysing finite-temperature cosmological phase transitions \cite{Wainwright:2011kj,Basler:2018cwe,Basler:2020nrq,Athron:2020sbe}.
In this approximation -- and in addition with the high-$T$ expansion that we assume -- the effective potential takes the form
\begin{align}
\label{eq:Veff_hack}
V^{\text{1-loop}}_{\text{thermal}} &= V_{\text{tree}} + V_{\text{CW}} + V_T + V_{\text{daisy}}, \nonumber \\
&= \underbrace{ V_{\text{tree}}  + \frac{M^2(v) T^2}{24} }_{V_{\text{thermal}}^{g^2}} - \underbrace{ T \frac{(M^2(v) + \Pi_T)^{\frac{3}{2}}}{12 \pi}}_{V_{\text{thermal}}^{g^3}} - \underbrace{\frac{M^4(v)}{(4\pi)^2} \frac{L_b(\mu)}{4}}_{\subset V_{\text{thermal}}^{g^4}} +\ \mr{const},
\end{align}
where, following Refs.~\cite{Farakos:1994kx,Kajantie:1995dw}, we have defined ($\gammaE$ is the Euler-Macheroni constant)
\begin{equation}
L_b(\mu) \equiv 2\log \left(\frac{e^{\gammaE } \rg }{4 \pi  T}\right).
\end{equation} 
Definitions of the various different terms in first line of Eq.~\eqref{eq:Veff_hack} are standard, and are given in full in Appendix~\ref{sec:details}, together with a derivation of the second line.
The second line utilises a split between soft and hard terms, originating respectively from the zero and non-zero Matsubara modes.
In this expression powers of $\gsq$ are separated but zero temperature and high temperature pieces are mixed.
The last term arises at one-loop, but is of order $\mc{O}(\gsqsq v^4)$, just as the Coleman-Weinberg potential at zero temperature.
However, due to the enhancements of infrared physics, this is not the full $\mathcal{O}(g^4)$ piece of the thermal potential, but only a part thereof, as we will see in next section.

In analogy with the result at zero temperature, it might be guessed that the renormalisation group running of this one-loop approximation, Eq.~\eqref{eq:Veff_hack}, is subdominant.
However, as we will show, this is not correct.
Explicitly for this scalar theory, the leading order running is
\begin{align}
\drg{} V_{\text{thermal}}^\rmi{1-loop} (v) &= 
\underbrace{\drg{} V_\rmi{tree} (v)+  \drg{}\Big(- \frac{M^4(v)}{(4\pi)^2} \frac{L_b(\mu)}{4}\Big)  }_{\rm cancels}
+ \underbrace{\drg{} \Big(\frac{M^2(v) T^2}{24}\Big)}_{\rm uncancelled}
\nonumber \\
&\quad 
-  \underbrace{\drg{} T \frac{(M^2(v) + \Pi_T)^{\frac{3}{2}}}{12 \pi} }_{\rm subdominant},
\label{eq:lo-thermal-scale-dependence}
\end{align}
where we have dropped irrelevant constant terms.
The scale dependence of the first two terms cancels exactly as in the zero temperature case since
\begin{align}
\drg{}\Big(- \frac{M^4}{(4\pi)^2} \frac{L_b(\mu)}{4}\Big)  = \drg{} V_{\text{CW}}.
\end{align}
However, at the same (leading) order there is the following leftover uncancelled term, which is (dropping an irrelevant constant)
\begin{align}
\drg{} V_{\text{thermal}}^\rmi{1-loop} (v) &= \drg{} \Big(\frac{M^2(v) T^2}{24}\Big), \\
 &= \frac{1}{2}\drg{\Pi_T} v^2. \label{eq:troublesome-term-again}
\end{align}
This is the troublesome term foreshadowed in Eq.~\eqref{eq:troublesome-term} in Sec.~\ref{sec:intro}.

The scale dependence does not cancel at leading order, in contrast to Eq.~\eqref{eq:zero-scale-dependence} at zero temperature.
As a consequence, at high temperature the running of couplings is not subdominant in this approximation, being of the same order as the Coleman-Weinberg term itself.
That is, at high temperatures the one-loop effective potential is scale dependent at order $\mc{O}(\gsqsq)$, whereas at zero temperature the scale dependence of the one-loop potential was only of order $\mc{O}( \gsqsqsq )$.
In fact, this scale dependence is the lowest order that scale dependence could arise,
being the same order as the running of the tree-level potential.

This equation,
\begin{align}
\drg{} V_{\text{thermal}}^\rmi{1-loop} (v) &= \frac{1}{2}\drg{\Pi_T} v^2,
\end{align}
also holds to leading order in more complicated theories.
In theories with non-$Z_2$-symmetric scalars, there is also a term at this order arising from thermal contributions to the tadpole; see for example Refs.~\cite{Gould:2021dzl,Schicho:2021gca,Niemi:2021qvp}.
For a beyond the Standard Model theory with an extra $Z_2$-symmetric scalar $S$ the thermal self-energy of the Higgs scalar is schematically
\begin{align}
\label{eq:BSM-thermal-mass}
\Pi^{\text{BSM}}_{T} = \frac{T^2}{16}\Big(3g^2 + g'^2 + 4 g^2_Y + 8 \lambda_h + \mathcal{C} \lambda_p \Big),
\end{align}
where $g,g'$ are SU(2) and U(1) gauge couplings, $g_Y$ is top Yukawa coupling (effect from other Yukawa couplings is numerically negligible), $\lambda_h$ the Higgs self-interaction and $\lambda_p $ the portal coupling between Higgs and scalar $S$.
The numerical factor $\mathcal{C}$ depends on the representation of $S$ under the gauge symmetries, 
for e.g. $\mathcal{C}= 2/3$ in the real-singlet extended SM~\cite{Schicho:2021gca},
$\mathcal{C}= 2$ in the real-triplet extended SM~\cite{Niemi:2018asa} 
and $\mathcal{C}= 4/3$ in the Two-Higgs Doublet Model (2HDM)~\cite{Gorda:2018hvi}.%
\footnote{In these references, the Higgs ($H$) portal interaction is defined as $\frac{1}{2}\lambda_p S^2 H^\dagger H$ for the singlet ($S$), $\frac{1}{2} \lambda_p \Sigma^a \Sigma^a  H^\dagger H$ for the triplet ($\Sigma^a$) and $\lambda_3 (H^\dagger H)(\phi^\dagger \phi) + \lambda_4 (H^\dagger \phi)(\phi^\dagger H)$ with $\lambda_p \rightarrow 2\lambda_3 + \lambda_4$ for the doublet ($\phi$). 
}

In the pure SM case, the leftover scale dependency is numerically dominated by the top quark since $g^2_Y > g^2, {g'}^2, \lambda_h$. The situation is even worse in those BSM theories where large portal couplings  $\lambda_p/(4\pi) \lesssim 1$ are required to change the character of the electroweak phase transition from a crossover to a strong first order transition.
The running of these large portal couplings prevents an accurate determination of the critical temperature (for example see Fig.~12 in Ref.~\cite{Kainulainen:2019kyp}), as well as all other thermodynamic parameters.

For the case of radiatively-induced first-order phase transitions, our conclusions regarding the magnitude of the renormalisation scale dependence are in fact an underestimate.
For the SM with light Higgs, this happens for $\lambda_h \sim g^3$, 
(see for example Ref.~\cite{Arnold:1992rz}, or for a recent discussion Ref.~\cite{Ekstedt:2020abj}),
and in the vicinity of the transition the $\mc{O}(g^2)$ terms in the effective potential partially cancel, leaving an $\mc{O}(g^3)$ remainder.
As a result the renormalisation scale dependence of Eq.~\eqref{eq:daisy} is of the same order as that of the tree-level potential, yielding an additional uncancelled scale dependence at leading order.

\section{Scale independence at high temperature}
\label{sec:rg-impr-high-T}

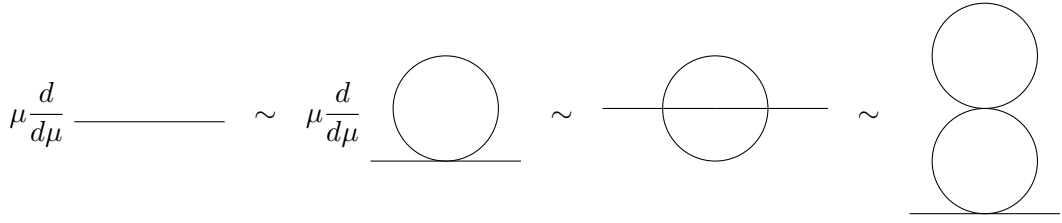
\begin{figure}[t]
    \centering
    \begin{align*}
    \mu\frac{d}{d\mu}\;
\begin{gathered}
	\begin{tikzpicture}[scale=1, transform shape]
    \begin{feynman}
    \treelevel{plain}
    \end{feynman}
    \end{tikzpicture}
\end{gathered}
\quad {\huge \sim} \quad
\mu\frac{d}{d\mu}\;
\begin{gathered}
	\begin{tikzpicture}[scale=1, transform shape]
    \begin{feynman}
    \selfenergyfour{plain}{plain}
    \end{feynman}
    \end{tikzpicture}
\end{gathered}
\quad {\huge \sim} \quad
\begin{gathered}
	\begin{tikzpicture}[scale=1, transform shape]
    \begin{feynman}
    \sunset{plain}{plain}
    \end{feynman}
    \end{tikzpicture}
\end{gathered}
\quad {\huge \sim} \quad
\begin{gathered}
	\begin{tikzpicture}[scale=1, transform shape]
    \begin{feynman}
    \cactus{plain}{plain}
    \end{feynman}
    \end{tikzpicture}
\end{gathered}
\end{align*}
    \caption{\label{fig:NLO} At high temperature, the running of tree-level parameters is the same order as the running of the one-loop thermal mass, which in turn is the same order as explicit logarithms of renormalisation scale at two-loop order.
    All these terms are of order $\mc{O}(g^4T^2)$.
    }
    \label{fig:rge-diagrams}
\end{figure}

The one-loop approximation to the thermal effective potential given in Eq.~\eqref{eq:Veff_hack} is incomplete at $\mc{O}(\gsqsq)$.
This is the reason for the residual renormalisation scale dependence at this order.
In the following, we will show explicitly how the scale dependence cancels in a complete calculation at $\mc{O}(\gsqsq)$. This requires the computation of two-loop Feynman diagrams, including two-loop corrections to the thermal mass, as illustrated schematically in Fig~\ref{fig:NLO}.

Perhaps the most widely used way to derive the full $\mc{O}(\gsqsq)$ thermal effective potential of a given theory is to utilise high-temperature dimensional reduction to a three-dimensional effective field theory (3d EFT);
see Refs.~\cite{Farakos:1994kx,Braaten:1995cm,Kajantie:1995dw,Andersen:1997zx} for the original literature and Refs.~\cite{Andersen:2004fp,Laine:2016hma,Croon:2020cgk,Gould:2021dzl,Schicho:2021gca} for reviews.
The technique of dimensional reduction is a systematic approach to the resummations required at high temperature, order-by-order in powers of couplings.
The $\mc{O}(\gsqsq)$ result for the $Z_2$-symmetric real scalar theory
-- that diagrammatically requires a two-loop determination --
has been derived long ago~\cite{Arnold:1992rz,Farakos:1994kx,Braaten:1995cm,Andersen:1997zx}.
Expanded to this order, the result reads
\begin{align}
\label{eq:veff-NNLO}
 V_{\text{thermal}}^{g^4}(v) &= \frac{1}{(4\pi)^2}\Bigg[
\frac{1}{12} \gsqsq T^2 v^2 \left(
\frac{1}{2}\log \left(\frac{M^2(v)+\Pi_T}{T ^2}\right)
-\frac{3}{8}L_b(\mu)
- c
+ \frac{1}{4}
\right)\Bigg]
\nonumber \\
&+\frac{1}{(4\pi)^2}\Big(
-\frac{1}{4} \gsq  m^2 v^2 L_b(\mu)
-\frac{1}{16} \gsqsq v^4 L_b(\mu)
\Big)
+ \mr{const},
\end{align}
where, following Refs.~\cite{Farakos:1994kx,Kajantie:1995dw}, we have introduced the notation
\begin{equation}
c \equiv -\log \left(\frac{3e^{\gammaE/2}A^6}{4 \pi} \right) = -0.348723\dots ,
\end{equation}
where $A$ is the Glaisher-Kinkelin constant and again we may use the freedom to add a constant to the potential to set $V_{\text{thermal}}(0)=0$.
The full result for the two-loop effective potential within the 3d EFT can be found in Eq.~\eqref{eq:3d-veff} in the appendix.%
\footnote{
Note that in typical computations that utilise dimensional reduction, one does not expand and truncate perturbative computations in terms of the 4d coupling expansion. In this way, higher order resummations are captured, and are kept to improve convergence; see thermal QCD Refs.~\cite{Blaizot:2003iq,Laine:2006cp}.
}

This result differs from the one-loop thermal effective potential, Eq.~\eqref{eq:Veff_hack}, by the terms on the first line of Eq.~\eqref{eq:veff-NNLO}, in square brackets.
From the perspective of high-temperature dimensional reduction, these terms arise from two sources: from the two-loop correction to the thermal mass, as well as from two-loop vacuum diagrams within the 3d EFT.
Despite arising from two-loop diagrams, the terms are clearly of the same size as the one-loop Coleman-Weinberg terms for $v\sim T$, both in terms of powers of $g^2$ and in terms of powers of $1/(4\pi)$.

This result for the $\mc{O}(\gsqsq)$ part of the effective potential depends explicitly on the renormalisation scale $\rg$, through the terms $L_b(\mu)$, as well as implicitly through the \MSbar\ parameters, though this latter dependence is of higher order.
The leading order renormalisation scale dependence is
\begin{equation}
\drg{}V_{\text{thermal}}^{g^4} (v) = -\frac{1}{(4\pi)^2}\left(
\frac{1}{2} \gsq  m^2 v^2
+ \frac{1}{16} \gsqsq T^2 v^2
+ \frac{1}{8} \gsqsq v^4\right).
\end{equation}
This exactly cancels the implicit renormalisation scale dependence of the leading order potential, Eq.~\eqref{eq:lo-thermal-scale-dependence}, so that
\begin{equation}
\label{eq:zero-thermal-scale-dependence}
\drg{} \left(V_{\text{thermal}}^{g^2}(v) + V_{\text{thermal}}^{g^3}(v) + V_{\text{thermal}}^{g^4}(v) \right) = 0.
\end{equation}
The cancellation however receives corrections which are of higher order.
The leading corrections are $\mathcal{O}(g^5)$ due to the (implicit) renormalisation scale dependence of the $\mathcal{O}(g^3)$ term.

Cancellations analogous to that of Eq.~\eqref{eq:zero-thermal-scale-dependence} happen generically, also in more complicated theories, as indeed they must.
This has been explicitly verified in the case of the SM in 
Refs.~{\cite{Arnold:1992rz,Buchmuller:1995sf,Farakos:1994kx,Kajantie:1995dw}. 
Similar computations in theories beyond the SM include \cite{Gorda:2018hvi,Kainulainen:2019kyp} (2HDM), \cite{Niemi:2018asa,Niemi:2020hto} (real-triplet) and \cite{Schicho:2021gca,Niemi:2021qvp} (real-singlet); works that all utilise dimensional reduction to three-dimensional effective theories at high temperature.
The cancellation of scale dependence is an important consistency check in the construction of these high temperature effective field theories.

In the computation of Eq.~\eqref{eq:veff-NNLO} both UV and IR logarithms arise, though in Eq.~\eqref{eq:veff-NNLO} the renormalisation scale dependence of the IR logarithms has already cancelled.
The UV logarithms are related to the usual renormalisation group running of couplings.
The latter, the IR logarithms, are instead related to the renormalisation group running within the 3d EFT~\cite{Farakos:1994kx}, as is characteristic of EFTs~\cite{Manohar:2018aog}.
Making use of this, the renormalisation scale of the IR logarithms can be replaced with a new scale, $\mu_3$.
The scale $\mu_3$ can then be set independently of $\mu$, thereby reducing the occurrence of large logarithms and improving the convergence of perturbation theory.
It also has the beneficial effect of separating out the effects of IR and UV scales, which we will make use of in the following.

\section{The consequences for equilibrium thermodynamics}
\label{sec:equilibrium}

The residual renormalisation scale dependence at $\mc{O}(g^4)$ of the common one-loop approximation to the thermal effective potential, Eq.~\eqref{eq:Veff_hack}, implies that there are corresponding theoretical uncertainties for all physical quantities computed in this approximation.
In this section, we focus on quantities which can be computed solely based on knowledge of the thermal effective potential.
These are necessarily bulk equilibrium quantities, and we focus on the critical temperature $T_c$ and the phase transition strength $\alpha$ evaluated at $T_c$.
This latter quantity is proportional to the latent heat when evaluated at $T_c$; see Ref.~\cite{Croon:2020cgk} for a definition.
We postpone to Sec.~\ref{sec:gw} discussion of the bubble nucleation rate, which is also necessary for determination of the gravitational wave spectrum, but which does not involve the effective potential in its computation.

As an explicit test of the magnitude of the residual renormalisation scale dependence, 
we investigate the simplest extension of the Standard Model with a first-order phase transition, namely, the Standard Model plus a real singlet scalar field (xSM).
The computations require generalising the results of earlier sections to include the Standard Model field content.
The details for this can be found in Refs.~\cite{Brauner:2016fla,Schicho:2021gca,Niemi:2021qvp}, where the high-temperature 3d EFT of this model was constructed and the effective potential computed to two-loop order.
The calculation of equilibrium quantities is thus complete at $\mc{O}(g^4)$.%
\footnote{
In fact, there are some contributions still missing at $\mc{O}(g^4)$, also left out in Ref.~\cite{Kajantie:1995dw}, which come from the  temporal components of gauge fields in the second step of dimensional reduction and which we expect to be numerically small.
One way to include these terms would be to perform the phase transition analysis at the soft scale, rather than integrating these fields out.
}

In computing $T_c$ and $\alpha_c \equiv \alpha(T_c)$ we follow the 3d~approach outlined in Ref.~\cite{Croon:2020cgk}.
That is, we perform dimensional reduction and then compute the relevant thermodynamic quantities in an $\hbar$-expansion within the 3d EFT, thereby ensuring order-by-order gauge invariance.
To circumvent the difficulties of phase transitions which are radiatively induced at the soft (or ultrasoft) scale, we focus on two benchmark points with a two-step transition.
The first step is a transition in the singlet direction, followed by a transition to the usual Higgs minimum.
For the second step of the transition, there is a tree-level barrier between phases, leading to a stronger transition.
We study only this second step.
The coupling expansion is more slowly convergent for radiatively induced one-step transitions, being essentially an expansion in $\sqrt{g}$~\cite{Arnold:1992rz,Ekstedt:2020abj}, rather than in $g$ as for the tree-level driven two-step transitions we focus on.

A parameter point in the $Z_2$-symmetric xSM can be specified by the values of $\{M_\sigma, \lambda_m, \lambda_\sigma \}$ where $M_\sigma$ is the pole mass of the singlet scalar, $\lambda_m$ is the Higgs-singlet portal coupling and $\lambda_\sigma$ is the singlet self-coupling~\cite{Schicho:2021gca}, 
the latter two are denoted as $a_2$ and $b_4$ respectively in Ref.~\cite{Niemi:2021qvp}.
We choose the following two 
illustrative benchmark points, BM1 and BM2
\begin{align} \label{eq:benchmark_parameters}
\text{BM1:} \quad \{M_\sigma, \lambda_m, \lambda_\sigma \} &= \{160~\mathrm{GeV}, 1.1, 0.45 \},\\ 
\text{BM2:} \quad \{M_\sigma, \lambda_m, \lambda_\sigma \} &= \{160~\mathrm{GeV}, 1.4, 1.4 \}.
\end{align}
The couplings are in the \MSbar\ scheme, at an input renormalisation scale equal to the mass of the $Z$ boson.
The one-loop relations between \MSbar\ parameters and physical observables are taken from Ref.~\cite{Niemi:2021qvp}.
Such $\mc{O}(1)$ couplings are typical of strong first-order phase transitions at the electroweak scale (see for example the benchmark points in Ref.~\cite{Caprini:2019egz}), 
making more acute the importance of resolving the issue of renormalisation scale dependence, stemming from missing $\mc{O}(g^4)$ corrections.

At these benchmark points we compare the following approximations, in decreasing order of accuracy:
\begin{enumerate}
\item[(i)] $\mc{O}(g^4)$: dimensional reduction at $\mc{O}(g^4)$, and $\mc{O}(\hbar^2)$ within EFT,
\item[(ii)] one-loop:
dimensional reduction at one-loop, and $\mc{O}(\hbar^1)$ within EFT,
\item[(iii)] $\mc{O}(g^3)$: dimensional reduction at $\mc{O}(g^3)$, and $\mc{O}(\hbar^1)$ within EFT,
\item[(iv)] $\mc{O}(g^2)$: dimensional reduction at $\mc{O}(g^2)$, and $\mc{O}(\hbar^0)$ within EFT.
\end{enumerate}
Note that the loop expansion of the effective potential within the EFT is an expansion in powers of $\hbar \sim g$, starting with the tree-level potential at $\mc{O}(g^2)$.
In the above the order at which dimensional reduction has been carried out is matched by the order of the loop-expansion within the EFT.

In (i)-(iv) above, we have separated out the orders at which the (UV) dimensional reduction was carried out, from those at which the (IR) calculation within the 3d EFT was carried out.
The former depends on the nonzero Matsubara modes, and the latter on the zero Matsubara modes.
We have done this to highlight how the calculation factorises into these two parts, and so too does the renormalisation scale dependence.
While the dimensional reduction step depends on the original renormalisation scale $\mu$, the calculation within the 3d EFT depends on its own renormalisation scale $\mu_3$~\cite{Farakos:1994kx}.

The one-loop approach (ii) is approximately equivalent to the usual one based on the one-loop daisy-resummed potential of Eq.~\eqref{eq:Veff_hack}~\cite{Croon:2020cgk}.
However, approach (ii) also includes the effects of one-loop wavefunction renormalisation, which enter through the dimensional reduction step, thus
it includes all logarithms of scale which are present at zero temperature, including the effects of nonzero anomalous dimension; see Eq.~\eqref{eq:rge_appendix}.
As a consequence, the magnitude of the renormalisation scale dependence of approach (ii) gives a conservative estimate for one-loop approaches in the literature, which typically neglect this effect.
Further, by utilising dimensional reduction, we are able to maintain order-by-order gauge invariance and also to avoid double counting problems and an uncontrolled derivative expansion in the bubble nucleation calculation in Sec.~\ref{sec:gw}.

\begin{figure}[t]
\begin{subfigure}{0.5\textwidth}
    \centering
    \includegraphics[width=\textwidth]{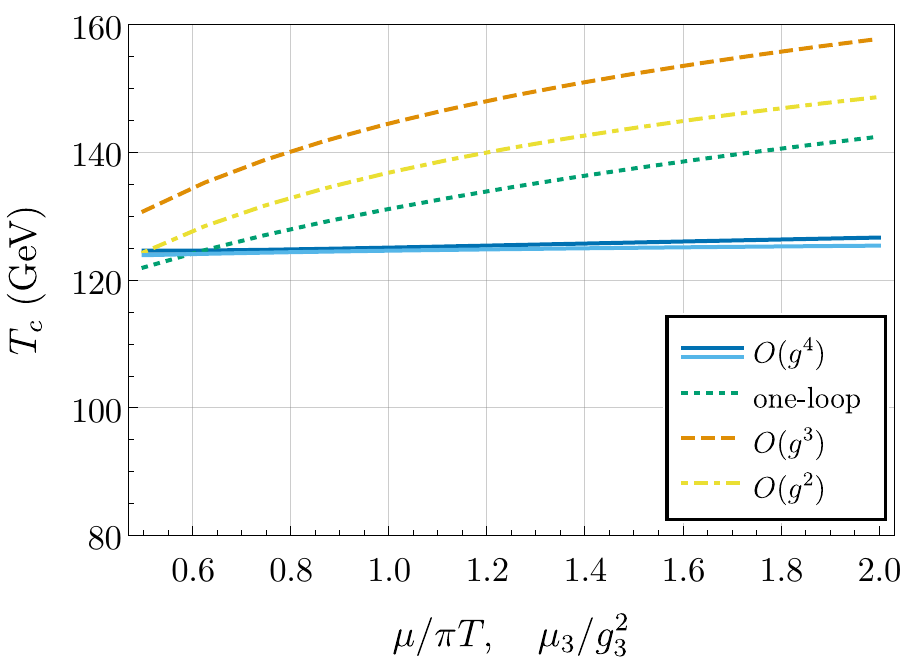}
    \label{fig:Tc_BM1}
\end{subfigure}
\begin{subfigure}{0.5\textwidth}
    \centering
    \includegraphics[width=\textwidth]{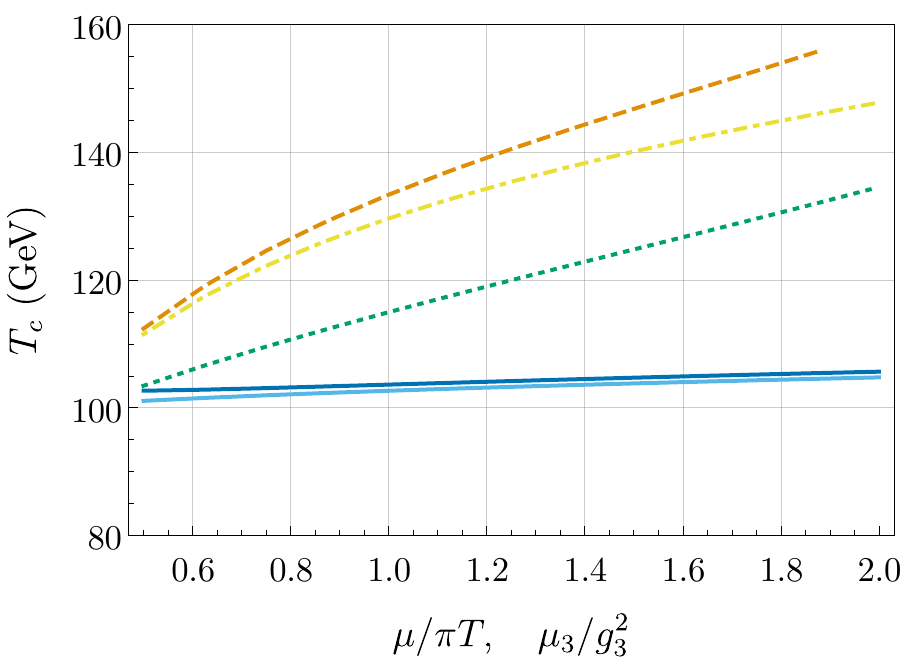}
    \label{fig:Tc_BM2}
\end{subfigure}
\\
\begin{subfigure}{0.5\textwidth}
    \centering
    \includegraphics[width=\textwidth]{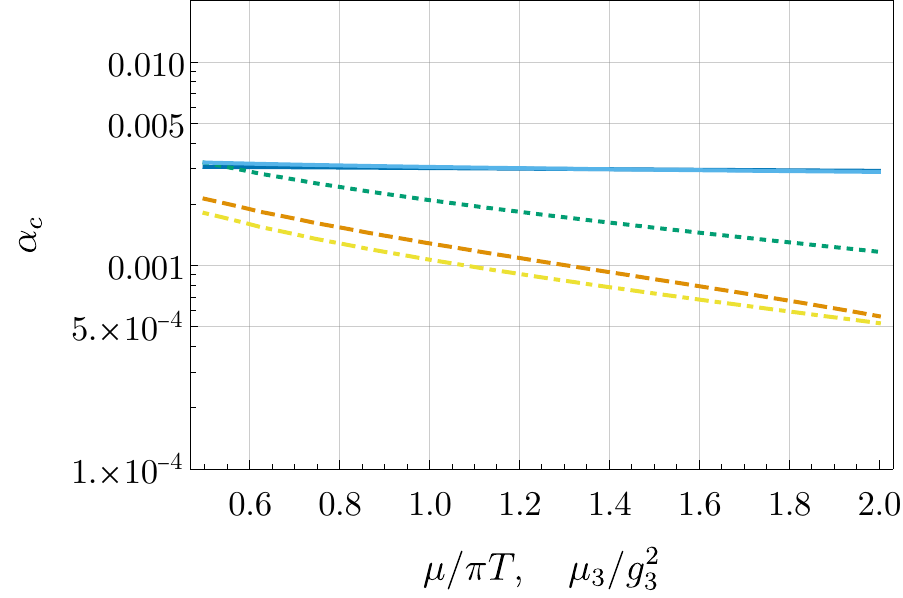}
    \label{fig:alphac_BM1}
\end{subfigure}
\begin{subfigure}{0.5\textwidth}
    \centering
    \includegraphics[width=\textwidth]{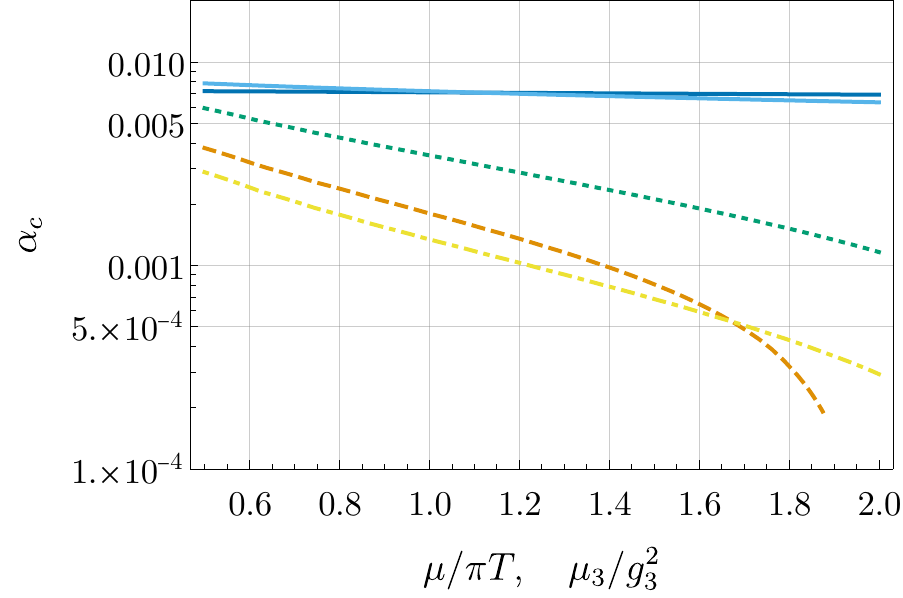}
    \label{fig:alphac_BM2}
\end{subfigure}
\caption{
$T_c$ and $\alpha_c$ as functions of the renormalisation scales, for BM1 on the left and BM2 on the right.
For the $\mc{O}(g^4)$ approximation, dark blue is $\mu$-dependence and light blue is $\mu_3$-dependence. 
Only the full 
$\mc{O}(g^4)$ 
accurate approximation shows signs of a controlled RG-scale dependence, 
in accord with the generic arguments of previous sections. 
}
\label{fig:equilibrium}
\end{figure}

In Fig.~\ref{fig:equilibrium} we plot $T_c$ and $\alpha_c$ as functions of the renormalisation scale $\mu$ in all four different approximations. In addition, for the $\mc{O}(g^4)$ approximation we show dependence on the 3d EFT renormalisation scale $\mu_3$. Note that in the 3d EFT running in terms of $\mu_3$ starts only at two-loop order, and hence we show it only for the $\mc{O}(g^4)$ approximation. On the x-axis, we use the dimensionless ratio $\mu_3/g^2_3$, where $g^2_3$ is the (dimensionfull) SU(2) gauge coupling in the 3d theory.  

Fig.~\ref{fig:equilibrium} shows that only in the full $\mc{O}(g^4)$ approximation is the renormalisation scale under control.
All other approximations show a striking dependence on renormalisation scale,
amounting to a few tens of percent for the critical temperature,
and a multiplicative factor of $\mc{O}(10)$ for $\alpha_c$.
This is especially marked for BM2, for which the couplings are larger and the transition stronger.
Comparing the one-loop approximation to the $\mc{O}(g^2)$ and $\mc{O}(g^3)$ approximations, one can see that the renormalisation scale dependence is somewhat smaller in the one-loop approximation but that it is nevertheless of the same order of magnitude, a numerical consequence of it being the same parametric order.
In the full $\mc{O}(g^4)$ approximation, the renormalisation scale dependence is at least an order of magnitude smaller, demonstrating numerically that its parametric order has indeed been reduced.
Dependence on $\mu_3$ is stronger than dependence on $\mu$, albeit not significantly.  

We conclude that for the equilibrium properties of the transition, the perturbative expansion is quantitatively under control only when all terms at $\mc{O}(g^4)$ are included, and this is not achieved at one-loop.
This conclusion follows naturally in light of earlier sections, as $\mc{O}(g^4)$ is the minimal order calculation that admits RG-improvement.

\section{The consequences for gravitational wave predictions}
\label{sec:gw}

The consequences 
of an incomplete perturbative computation at $\mc{O}(g^4)$ are particularly severe
for the gravitational wave spectrum produced by a first-order phase transition.
This was demonstrated in Ref.~\cite{Croon:2020cgk}, where it was found that missing $\mc{O}(g^4)$ terms, revealed through renormalisation scale dependence, give potentially the largest source of error for one-loop computations.%
\footnote{
See also Ref.~\cite{Guo:2021qcq} where the magnitude of various uncertainties~\cite{Jinno:2017ixd,Cutting:2019zws,Guo:2020grp,Giese:2020rtr,Giese:2020znk} due to macroscopic, rather than quantum field theoretic, physics were investigated. While certainly important, none of these are quite as severe as those we find here. Our approach to the macroscopic physics can be described as {\em moderately diligent} in the language of Ref.~\cite{Guo:2021qcq}.
}
The amplitude of the gravitational wave spectrum depends very sensitively on the thermodynamic parameters, and these in turn have relatively large uncertainties.
Although the theoretical arguments in Ref.~\cite{Croon:2020cgk} were general, the numerical study was restricted to a simplified version of the Standard Model Effective Field Theory.
The arguments of this article at hand pinpoint where the renormalisation scale dependence arises.
From this we expect the conclusions of Ref.~\cite{Croon:2020cgk} in this respect to hold rather generally.

In the previous section, we have presented four approximations
with which we have computed the purely equilibrium quantities $T_c$ and $\alpha_c$.
The gravitational wave spectrum depends additionally on the bubble nucleation rate, and the speed of the bubble wall growth, both of which are inherently real-time quantities, depending on the evolution of inhomogeneous field configurations.
As a consequence, they are significantly more challenging to compute.
In this article we have focused on equilibrium physics, and we refrain from overstepping this remit.
Fortunately, at leading exponential order the calculation of the bubble nucleation rate reduces to a purely equilibrium calculation~\cite{Langer:1969bc,langer1974metastable}, $\mc{O}(\hbar^0)$ within the 3d EFT.
For this we follow the approach outlined in Ref.~\cite{Croon:2020cgk}, though approximating the nucleation prefactor simply by $T^4$.
At $\mc{O}(\hbar^1)$ real-time physics enters the bubble nucleation rate (through the {\em dynamical prefactor}), as it does for the bubble wall speed at leading order.

As a consequence, we are not able to carry out a full $\mc{O}(g^4)$ calculation of the gravitational wave spectrum.
We can nevertheless compare different approximations to the dimensional reduction, though in each case the bubble nucleation calculation is carried out at $\mc{O}(\hbar^0)$ within the 3d EFT.
As far as we are aware, there does not exist a complete calculation of the thermal bubble nucleation rate at $\mc{O}(\hbar^1)$ for any quantum field theory,
yet it is necessary to extend to $\mc{O}(\hbar^2)$ in order to match the accuracy of the equilibrium calculations.
For the bubble wall speed we simply take $v_w=1$ in the gravitational wave amplitude, noting that there remains considerable debate in the literature as to the leading effects which contribute to this quantity; see for example Refs.~\cite{Bodeker:2017cim,Mancha:2020fzw,Mou:2020zyy,Hoeche:2020rsg,Laurent:2020gpg,Vanvlasselaer:2020niz}.
In the following, for simplicity, we will continue to use the same names for the approximations as in Sec.~\ref{sec:equilibrium}, though one should bear in mind these additional limitations with regard to bubble nucleation and the bubble wall speed.

The issue of renormalisation scale dependence is therefore more complicated for our calculations of the non-equilibrium quantities related to bubble nucleation.
As before, the calculation factorises into a UV part (dimensional reduction), and an IR part (within the 3d EFT), the latter only carried out at $\mc{O}(\hbar^0)$.
By carrying out the dimensional reduction at $\mc{O}(g^4)$, the $\mu$-dependence cancels, as it did for $T_c$ and $\alpha_c$.
Again this leads to significant improvements over the lower-order calculations.
However, the $\mu_3$-dependence arising from the IR does not cancel, as this would require an $\mc{O}(\hbar^2)$ calculation of the bubble nucleation rate.
This problematic $\mu_3$-dependence of the nucleation rate infects the GW signal, even if the equilibrium analysis is performed at $\mc{O}(g^4)$.

\begin{figure}[t]
\begin{subfigure}{0.5\textwidth}
    \centering
    \includegraphics[width=\textwidth]{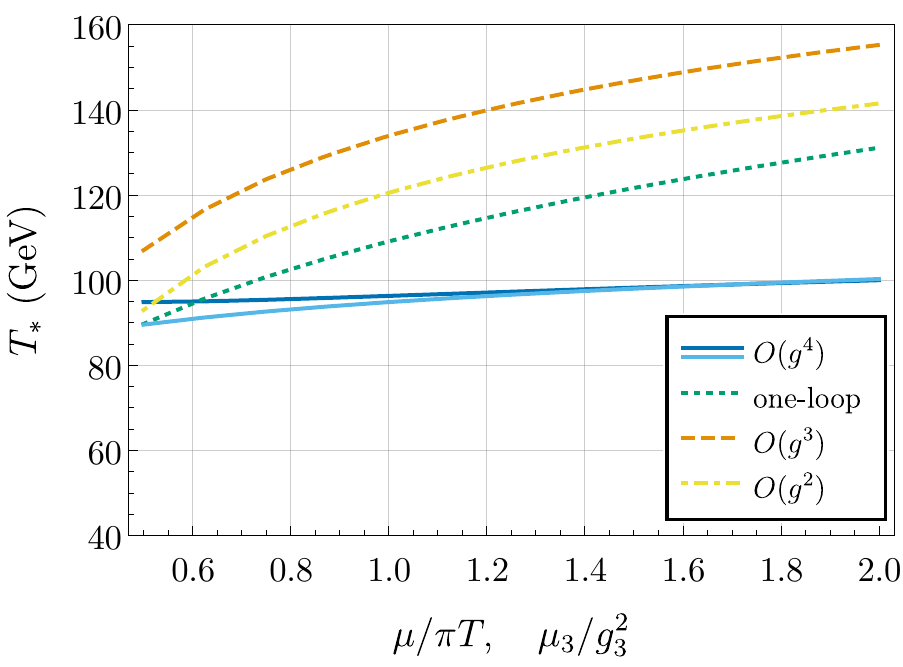}
    \label{fig:Tstar_BM1}
\end{subfigure}
\begin{subfigure}{0.5\textwidth}
    \centering
    \includegraphics[width=\textwidth]{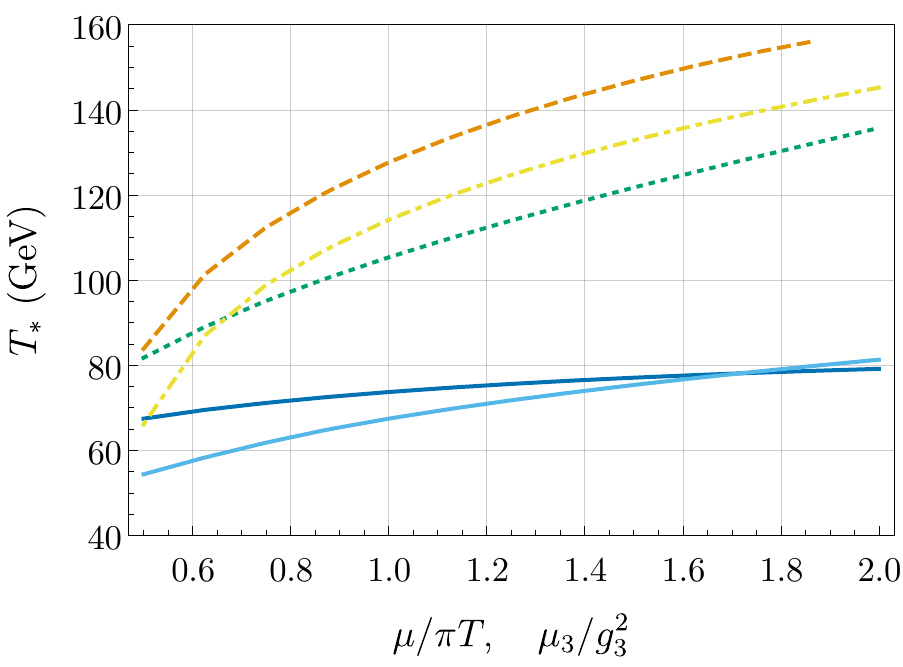}
    \label{fig:Tstar_BM2}
\end{subfigure}
\\
\begin{subfigure}{0.5\textwidth}
    \centering
    \includegraphics[width=\textwidth]{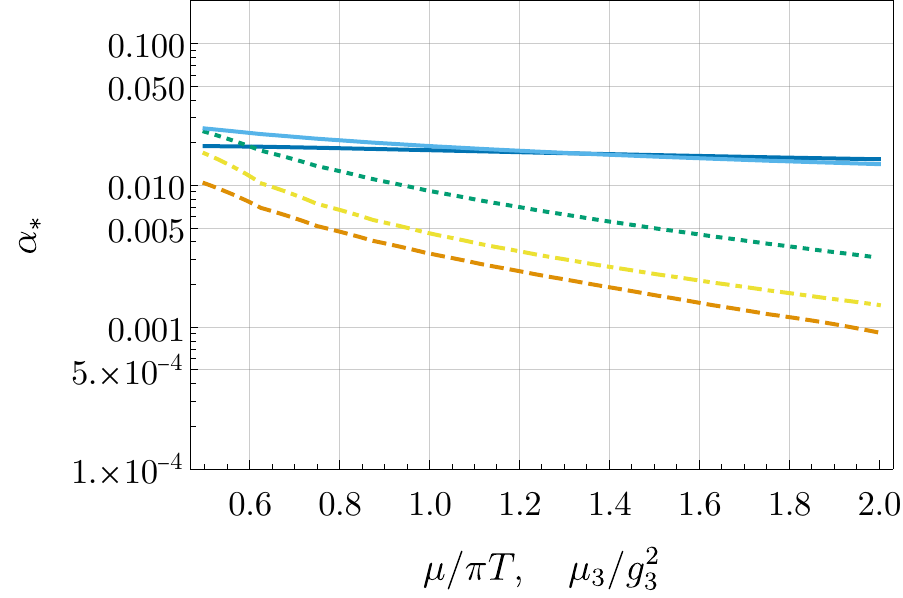}
    \label{fig:alphastar_BM1}
\end{subfigure}
\begin{subfigure}{0.5\textwidth}
    \centering
    \includegraphics[width=\textwidth]{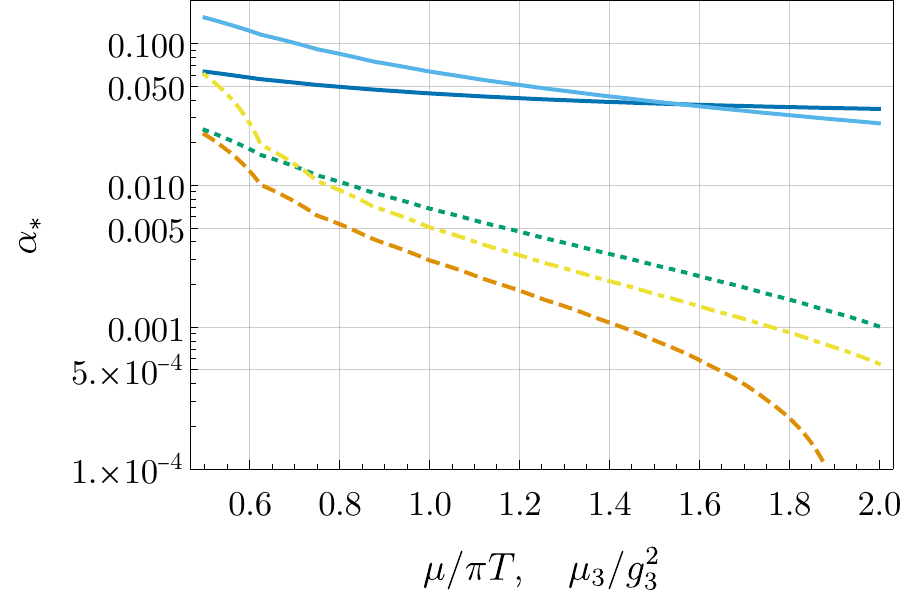}
    \label{fig:alphastar_BM2}
\end{subfigure}
\caption{
As Fig.~\ref{fig:equilibrium} but for $T_*$ and $\alpha_*$. The $\mc{O}(g^4)$ approximation still shows a weaker RG-scale dependence than the other approximations. However, compared to the purely equilibrium quantities in Fig.~\ref{fig:equilibrium}, dependence on $\mu_3$ is significantly stronger due to the limited $\mc{O}(\hbar^0)$ accuracy of the bubble nucleation calculation. This is especially true for BM2, for which the scalar couplings are larger.
}
\label{fig:Tstar_alphastar}
\end{figure}

\begin{figure}[t]
\begin{subfigure}{0.5\textwidth}
    \centering
    \includegraphics[width=\textwidth]{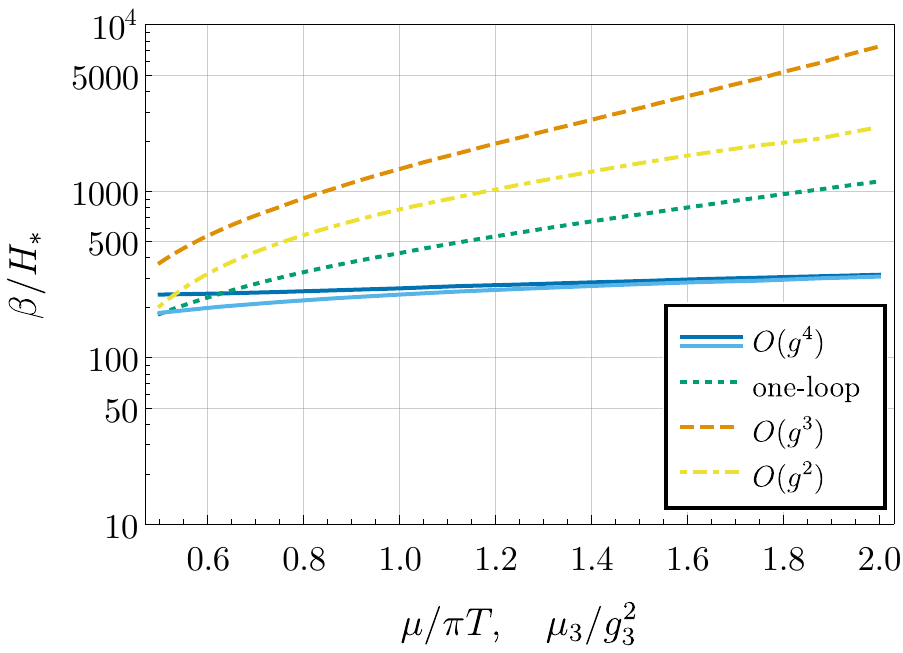}
    \label{fig:beta_BM1}
\end{subfigure}
\begin{subfigure}{0.5\textwidth}
    \centering
    \includegraphics[width=\textwidth]{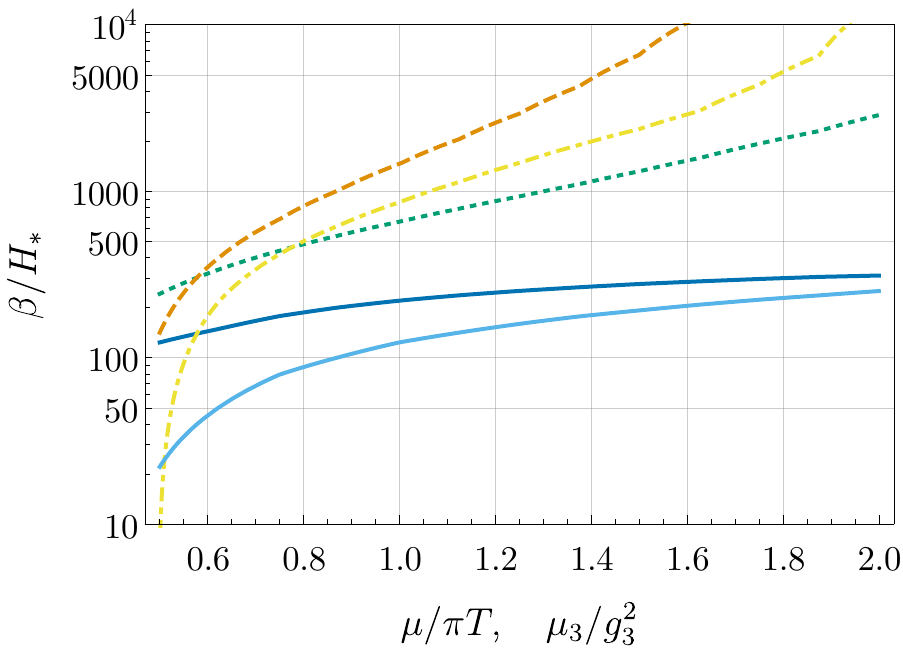}
    \label{fig:beta_BM2}
\end{subfigure}
\\
\begin{subfigure}{0.5\textwidth}
    \centering
    \includegraphics[width=\textwidth]{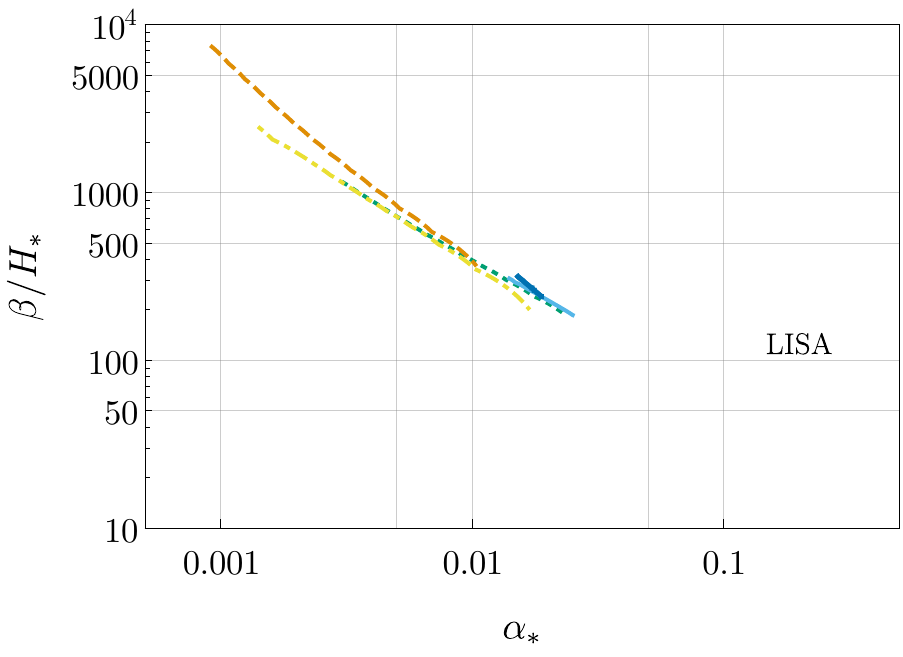}
    \label{fig:alphabeta_BM1}
\end{subfigure}
\begin{subfigure}{0.5\textwidth}
    \centering
    \includegraphics[width=\textwidth]{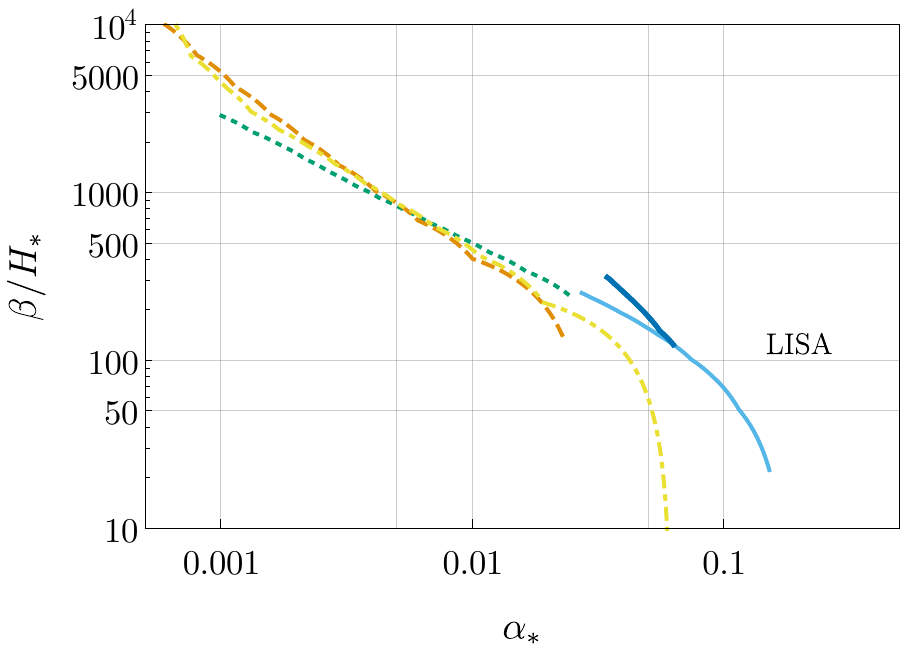}
    \label{fig:alphabeta_BM2}
\end{subfigure}
\caption{
Upper plots are as Fig.~\ref{fig:Tstar_alphastar} but for $\beta/H_*$. The lower plots show a revisualisation of the renormalisation scale dependence in the $(\alpha_*,\beta/H_*)$-plane. For context, the approximate region to which LISA is sensitive is indicated.
}
\label{fig:beta-alphabeta}
\end{figure}

In Figs.~\ref{fig:Tstar_alphastar} and \ref{fig:beta-alphabeta} we present our results at BM1 and BM2 for the thermodynamic parameters that go into the computation of the gravitational wave spectrum; see Ref.~\cite{Caprini:2019egz} for definitions. In Fig.~\ref{fig:Tstar_alphastar} we plot the percolation temperature ($T_*$) and the phase transition strength $\alpha_*\equiv \alpha(T_*)$, as functions of the RG scales.
For these, we show all four of the aforementioned approximations, similarly to Fig.~\ref{fig:equilibrium}.
In the $\mc{O}(g^3)$ approximation, the full range of RG-scale cannot be shown,
since for some values the second step of the phase transition disappears altogether.
Just as in Fig.~\ref{fig:equilibrium}, the $\mc{O}(g^2)$, $\mc{O}(g^3)$ and one-loop approximations show a strong RG-scale dependence, of the same order of magnitude for all three.
The approximation using $\mc{O}(g^4)$ dimensional reduction shows a markedly weaker renormalisation scale dependence than the other three approximations, especially regarding $\mu$.
However, for $T_*$ and $\alpha_*$ the $\mu_3$-dependence is much worse than it was for $T_c$ and $\alpha_c$, especially at BM2 which has larger scalar couplings.
As discussed, this is due to the limited $\mc{O}(\hbar^0)$ accuracy of the nucleation calculation.

In Fig.~\ref{fig:beta-alphabeta}, we show the inverse duration of the transition ($\beta/H_*$) as a function of the RG scales.
In addition, we replot $\beta/H_*$ against $\alpha_*$ in Fig.~\ref{fig:Tstar_alphastar}, emulating the plots produced in {\tt PTPlot}~\cite{Caprini:2019egz}. 
This revisualisation makes especially apparent the magnitude of the intrinsic uncertainty, for single benchmark points.
For all three of the approximations with less than the full $\mc{O}(g^4)$ dimensional reduction, the intrinsic uncertainty is strikingly large.
In particular for BM2, even in the best of our approximations, the variance in the $(\alpha_*,\beta/H_*)$-plane is still very large. 

Finally, using these thermodynamic parameters we plot the
gravitational wave spectrum due to sound waves~\cite{Hindmarsh:2017gnf,Caprini:2019egz}.
The gravitational wave signal of BM1
was already shown in Fig.~\ref{fig:gws-BM1} in Sec.~\ref{sec:intro}, together with the LISA Science Requirements sensitivity curve~\cite{LISAScienceRequirements}.
A similar plot for BM2 is given in Fig.~\ref{fig:gws-BM2}.
For simplicity, in these figures we only show the $\mc{O}(g^4)$ and one-loop approximations; the other approximations have larger uncertainties, shown in Table~\ref{tab:results}.
At BM1 the uncertainty band of the one-loop approximation spans over four orders of magnitude for the peak amplitude, making the prediction quite ambiguous.
In the $\mc{O}(g^4)$ approximation, the uncertainty is significantly smaller, but is still one order of magnitude for the peak amplitude. 
We note that upgrading the one-loop to the $\mc{O}(g^4)$ approximation leads to an increase in the peak amplitude, and shifts the location of the peak to smaller frequencies -- though these are expected to be model, and parameter point, specific trends.
However, based on our argumentation of the previous sections, the reduction of RG-scale dependence can be expected to hold generically.
For BM2, at which there are larger scalar couplings, the theoretical uncertainty in the gravitational wave signal is much larger, and as a prediction for experiments such as LISA it is very ambiguous.
In the one-loop approximation the peak amplitude varies by six orders of magnitude, and even in the $\mc{O}(g^4)$ approximation it varies by four orders of magnitude, chiefly a consequence of the uncertainty in the bubble nucleation rate.
From this we can conclude that, in order to pursue quantitatively accurate predictions, future work on the bubble nucleation rate is necessary. 

\begin{figure}[t]
    \centering
    \includegraphics[width=0.65\textwidth]{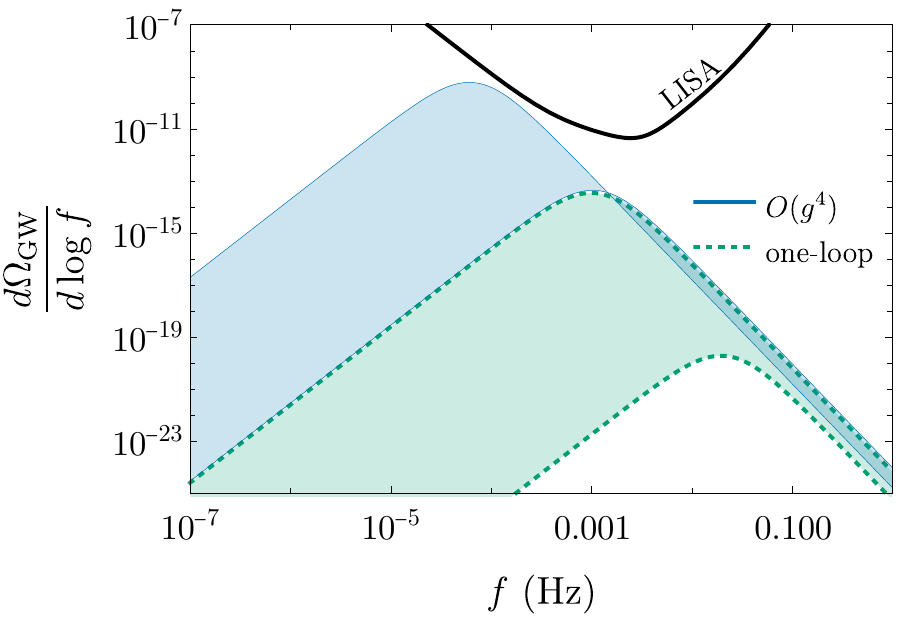}
    \caption{
    Similar to Fig.~\ref{fig:gws-BM1}, for BM2. Note that in both approximations the bubble nucleation rate is only accurate to $\mc{O}(\hbar^0)$ within the 3d EFT.
         The corresponding LISA signal-to-noise ratios for a 3 year mission profile~\cite{Caprini:2019egz} vary from 0.16 to 12 in the $\mc{O}(g^4)$ approximation, and from $1.5\times 10^{-9}$ to 0.12 in the one-loop approximation.
    }
    \label{fig:gws-BM2}
\end{figure}

\section{Discussion}
\label{sec:discussion}

\subsection*{Summary}

The main conclusions of this article can be restated rather simply.
As is well known, renormalisation scale invariance implies that, in the coupling expansion of the effective potential, lower order terms are linked by the implicit running of couplings to explicit logarithms of the renormalisation scale in higher order terms.
At zero temperature, the coupling expansion of the effective potential takes the form 
\begin{align} \label{eq:VVacuumCouplingExpansion}
V^{T=0}_{\text{eff}} = A_2 g^2 + A_4 g^4 + A_6 g^6 + \ldots \;.
\end{align}
This schematic formula presents a formal coupling expansion; in theories with multiple couplings there are expansions for each coupling, though in practice one typically establishes formal power counting rules for bookkeeping.
At zero temperature the coupling expansion and the loop expansion coincide;
$A_2$ is the tree-level term, $A_4$ arises purely at one-loop, and in general $n$-loop diagrams give the $A_{2(n+1)} g^{2(n+1)}$ term.
The running of couplings takes the form 
\begin{align}
\mu\frac{d g^2}{d\mu} = B_4 g^4 + B_6 g^6 + \ldots \;,
\end{align}
where $B_4$ is the one-loop term, $B_6$ is the two-loop term and so on.
The one-loop running of $A_2 g^2$ is an $\mc{O}(g^4)$ effect and is cancelled exactly by explicit logarithms in $A_4 g^4$, meaning that a one-loop calculation is sufficient to achieve renormalisation scale invariance at $\mc{O}(g^4)$, receiving corrections at $\mc{O}(g^6)$.
Explicitly, the cancellation takes the form
\begin{align}
\mu\frac{d V^{T=0}_{\text{eff}}}{d\mu}
&= \underbrace{\left(A_2 B_4 + \mu\frac{d A_4}{d\mu}\right)}_{\text{cancels}}g^4 + \mc{O}(g^6).
\end{align}

At high temperature, the enhancement of IR bosonic modes modifies the coupling expansion of the effective potential,
\begin{align} \label{eq:VThermalCouplingExpansion}
V^{\text{high-}T}_{\text{eff}} = a_2 g^2 + a_3 g^3 + a_4 g^4 + a_5 g^5 + \ldots \;.
\end{align}
Odd powers of $g$ arise, and loop orders are mixed in the expansion coefficients.
The coefficient $a_2$ receives contributions at both tree-level and one-loop, the coefficient $a_3$ comes from the (daisy-resummed) one-loop term, and the coefficient $a_4$ receives contributions at both (resummed) one- and two-loop.
However, thermal effects do not modify the runnings of couplings, as this depends purely on the UV.
So, one-loop running links the coefficients $a_i$ with $a_{i+2}$ (and higher loop running with $a_{i+4}$ and so forth),
\begin{align} 
\mu\frac{d V^{\text{high-}T}_{\text{eff}}}{d\mu} 
&= \underbrace{\left(a_2 B_4 + \mu\frac{d a_4}{d\mu}\right)}_{\text{cancels}}g^4 + \underbrace{\left(\frac{3}{2}a_3 B_4 + \mu\frac{d a_5}{d\mu}\right)}_{\text{cancels}}g^5 + \mc{O}(g^6).
\end{align}
The leading order running starts at $\mc{O}(g^4)$, and hence this is also the minimal accuracy for any cancellation of the renormalisation scale.
As $a_4$ receives contributions at (resummed) two-loop, no (resummed) one-loop calculation can achieve renormalisation group improvement at high temperatures.
Further, to achieve a residual uncertainty of $\mc{O}(g^6)$, equivalent to a one-loop calculation at zero temperature, requires also computing $a_5$, which receives IR contributions at three-loop order.

\subsection*{Outlook}

At two benchmark points in the xSM, we have demonstrated the numerical importance of these theoretical considerations for the determination of properties of the phase transition.
We have found that calculations at less than $\mc{O}(g^4)$ accuracy show a very large renormalisation scale dependence, in agreement with previous studies in other models~\cite{Kainulainen:2019kyp,Carena:2019une,Croon:2020cgk,Papaefstathiou:2020iag}.
We also remarked that while we were able to perform a complete $\mc{O}(g^4)$ calculation for $T_c$ and $\alpha_c$, this is currently out of reach for the bubble nucleation rate and bubble wall speed, imparting a limiting source of uncertainty for the gravitational wave spectrum.
Table~\ref{tab:results} presents a summary of the magnitude of the scale dependence in the four different approximations we adopted.

Though our numerical analysis was carried out merely at two benchmark points, the underlying arguments of Eqs.~\eqref{eq:VVacuumCouplingExpansion} and \eqref{eq:VThermalCouplingExpansion} are generic features of perturbative quantum field theory at zero and high temperature.
Thus, we expect the qualitative conclusions to apply to the wide variety of models of cosmological phase transitions considered in the literature.

\begin{table}[t]
\centering
\subfloat[BM1]{\begin{tabular}{|c|c|c|c|c|}
\hline
 & $\Delta T_c/T_c$ & $\Delta T_*/T_*$ & $\Delta \Omega/\Omega$ & $\Delta \mr{SNR}/\mr{SNR}$\\
\hline
$\mc{O}(g^4)$ & 0.02 & 0.12 & $1.4\times 10^1$ & $1.0\times 10^1$ \\ 
\hline
one-loop & 0.16 & 0.46 & $1.6\times 10^4$ & $7.3\times 10^4$ \\ 
\hline 
$\mc{O}(g^3)$ & 0.20 & 0.45 & $1.7\times 10^6$ & $5.0\times 10^8$ \\ 
\hline
$\mc{O}(g^2)$ & 0.19 & 0.52 & $2.1\times 10^5$ & $6.9\times 10^6$ \\
\hline
\end{tabular}}
\vspace{2mm}
\subfloat[BM2]{\begin{tabular}{|c|c|c|c|c|}
\hline
 & $\Delta T_c/T_c$ & $\Delta T_*/T_*$ & $\Delta \Omega/\Omega$ & $\Delta \mr{SNR}/\mr{SNR}$ \\
\hline
$\mc{O}(g^4)$ & 0.04 & 0.5 & $1.4\times 10^4$ & $7.7\times 10^1$\\ 
\hline
one-loop & 0.25 & 0.7 & $1.8\times 10^6$ & $8.3\times 10^7$\\ 
\hline
$\mc{O}(g^3)$%
\tablefootnote{For BM2 in the $\mc{O}(g^3)$ approximation, the existence of the phase transition is sensitive to the renormalisation scale over this range, so the values presented are instead obtained as the renormalisation scale varies by a slightly smaller factor of $3.7$.}
& 0.34 & 0.9 & $4.3\times 10^{10}$ & $5.6\times 10^{18}$\\ 
\hline
$\mc{O}(g^2)$ & 0.32 & 1.2 & $5.8\times 10^{11}$ & $9.1\times 10^{11}$\\
\hline
\end{tabular}}
	\caption{
    Summary of theoretical uncertainties due to residual renormalisation scale dependence for the four different approximations.
    In each case $\Delta X/X \equiv (\mr{max}(X)-\mr{min}(X))/\mr{min}(X)$, where the set $X$ consists of the values obtained as the renormalisation scale varies by a factor of 4.
    Here $\Omega$ refers to the gravitational wave peak amplitude and SNR to the LISA signal-to-noise ratio for a 3 year mission, computed using {\tt PTPlot}~\cite{Caprini:2019egz}.
    }
    \label{tab:results}
\end{table}

We thus infer that missing $\mc{O}(g^4)$ corrections to the thermal effective potential (and other thermodynamic quantities) are a crucially important, and likely a limiting theoretical uncertainty in predictions of the gravitational wave signal of cosmological phase transitions.
Their inclusion is therefore a necessary hurdle to surmount, if one wants to make quantitatively reliable predictions.
This applies also to electroweak baryogenesis, and other phenomena which depend on the dynamics of a cosmological first-order phase transition.
We advocate that the magnitude of the theoretical uncertainty should be acknowledged and assessed in future analyses of cosmological phase transitions.

We remark that while our $\mc{O}(g^4)$ calculation shows significant improvements over the lower-order calculations, the theoretical uncertainty in the GW peak amplitude is still rather large: one order of magnitude at BM1 and four orders of magnitude at BM2.
This motivates future work on the subject, with the aim of improving predictions further.
In particular, we have identified the bubble nucleation rate as likely a limiting source of theoretical uncertainty, once $\mc{O}(g^4)$ corrections are included for the equilibrium quantities.
This can be inferred from a comparison of Figs.~\ref{fig:equilibrium} and \ref{fig:Tstar_alphastar}, and is due to the nucleation rate being calculated at lower order $\mc{O}(\hbar^0)$ than the purely equilibrium quantities.
Even a complete calculation of the $\mc{O}(\hbar)$ corrections would be a first, and a significant advancement, whereas one must go to $\mc{O}(\hbar^2)$ in order to achieve parity with the equilibrium computations.%
\footnote{
Note that merely using the one- or two-loop effective potential for the calculation of the critical bubble would not constitute a genuine improvement, as this double-counts degrees of freedom in the path integral, and in turn relies upon an uncontrolled derivative expansion; see Ref.~\cite{Croon:2020cgk} Sec.~3.5 and references therein.
}

Nevertheless, higher order computations of equilibrium quantities would allow one to really test the convergence of perturbation theory, and the $\mc{O}(g^5)$ correction in particular may be sizeable, as was found for QCD~\cite{Zhai:1995ac,Braaten:1995jr}.
The $\mc{O}(g^5)$ correction was also found to significantly improve agreement between lattice and perturbation theory for relatively weak phase transitions in the real scalar theory~\cite{Gould:2021dzl}.
Computations of equilibrium thermodynamics at $\mc{O}(g^5)$~\cite{Zhai:1995ac,Braaten:1995jr,Gynther:2005dj,Laine:2015kra,Gould:2021dzl} and at even higher orders~\cite{Kajantie:2002wa,Gynther:2007bw,Andersen:2009ct} have been performed for several theories.
The underlying methodology for such higher-order calculations applies also to BSM electroweak phase transitions, though it nevertheless presents challenges.

It is important to emphasise that the discussion in this article is in the context of purely perturbative studies of thermal phase transitions.
In fact, due to Linde's Infrared Problem~\cite{Linde:1980ts} infinitely many (resummed) loop orders contribute to the effective potential at $\mc{O}(g^6)$, rendering perturbative approaches inherently incomplete.
Perhaps the only tool which can overcome this problem is lattice Monte-Carlo simulations~\cite{Farakos:1994xh,Kajantie:1995kf,Kajantie:1997hn,Laine:2000rm,Laine:2012jy,Kainulainen:2019kyp,Niemi:2020hto,Gould:2021dzl}.
Nevertheless, perturbation theory is a valuable guide, and one we can test our confidence in by comparison to the results of lattice simulations.

It may be argued, contrary to our conclusions, that in the context of BSM theories with unknown input parameters, precise calculations are unnecessary as the difference between different approximations may well be accommodated by a shift in the input parameters.
To this perspective we present the following counterarguments:
First, several orders of magnitude intrinsic uncertainty (as well as uncertainty regarding the order of the transition) is unsatisfactory on purely theoretical grounds, and naturally leads one to doubt the reliability of the calculations, and to aim to improve them where possible.
Second, even in full parameter scans, the range of possible gravitational wave signals which can be produced by a given model depends relatively sensitively on e.g.\ the functional form of the thermal effective potential~\cite{Kehayias:2009tn}, and hence on the perturbative treatment.
Third, by the time the second generation of gravitational wave detectors, such as LISA and Taiji, are due for launch, the LHC Runs 3, 4 and part of Run 5 will have been completed, as well as many other experiments searching for BSM physics.
Thus, in the future scenario whereby particle physics experiments are able to point towards some specific type of BSM theory, or -- being even more optimistic -- even a narrow region of its parameter space, we want to be in a position to compute the thermodynamics relevant for GW predictions as accurately as possible, and for this it is important to acknowledge that accuracy less than $\mc{O}(g^4)$ is potentially unreliable.
Finally, reversing this argument, if a stochastic GW background of primordial origin is observed, it is crucial to be able to make as accurate predictions as possible, for there to be any hope to reverse-engineer the underlying particle physics model -- i.e.\ the LISA inverse problem~\cite{Croon:2018erz}.

%
\section*{Acknowledgements}
The authors wish to thank
Lauri Niemi,
Philipp Schicho and
Juuso {\"O}sterman
for their work in collaboration which provided the foundations for our xSM calculations.
We would also like to thank
Djuna Croon,
Joonas Hirvonen,
Thomas Konstandin,
Michael J.~Ramsey-Musolf,
Jorinde van de Vis,
David Weir and
Graham White
for enlightening discussions.
O.G.\ was supported by U.K.~Science and Technology Facilities Council (STFC) Consolidated grant ST/T000732/1.

%
\appendix

\numberwithin{equation}{section}

\renewcommand{\thesection}{\Alph{section}}
\renewcommand{\thesubsection}{\Alph{section}.\arabic{subsection}}
\renewcommand{\theequation}{\Alph{section}.\arabic{equation}}


\section{Appendix}
\label{sec:details}

\paragraph{Renormalisation and quantum corrections}

For a review of the zero temperature effective potential in electroweak theories, see Ref.~\cite{Sher:1988mj}.
The parameters and the field appearing in the Lagrangian density in Eq.~\eqref{eq:lagrangian} are strictly speaking bare quantities, 
related to their renormalised counterparts as
\begin{align}
\phi_{(b)} &\equiv Z^{\frac{1}{2}} \phi = (1+\delta Z)^{\frac{1}{2}} \phi, \\
m^2_{(b)} &\equiv Z^{-1}(m^2 + \delta m^2), \\
\gsq_{(b)} &\equiv Z^{-2} \rg^{2\epsilon} (\gsq + \delta \gsq),
\end{align}
where renormalised parameters are denoted without subscripts.
Here, the wave function renormalisation $Z=1$ and $\delta Z = 0$ as there are no divergent topologies with external momentum dependence at one-loop order.
For the other counterterms, we define at one-loop order
\begin{align}
\delta m^2 &= \frac{1}{(4\pi)^2 \epsilon}  \frac{1}{2} \gsq m^2 , \quad \quad
\delta \gsq = \frac{1}{(4\pi)^2 \epsilon}  \frac{3}{2} \gsqsq. 
\end{align}
We ignore the vacuum counterterm required to cancel the field independent divergence of the effective potential.
Bare parameters are independent of the renormalisation scale, and hence we can solve for the scale dependence or running of the renormalised parameters%
\footnote{
Note that the scale dependence of bare parameters is required to vanish order-by-order in $\epsilon$ and $g^2$, and hence $\beta(\gsq)$ starts at order $\epsilon g^2$.
}
\begin{align}
\rg \frac{d}{d\rg} \gsq_{(b)} &= 0 \implies \beta(\gsq) \equiv \rg \frac{d}{d\rg} \gsq = 2 \epsilon \Big(-\gsq + \delta \gsq \Big), \\
\rg \frac{d}{d\rg} m^2_{(b)} &= 0 \implies \beta(m^2) \equiv  \rg \frac{d}{d\rg} m^2 = -\frac{1}{(4\pi)^2\epsilon} \Big( \beta(\gsq) m^2 + \mathcal{O}(\gsqsq)  \Big),
\end{align}
In the limit $\epsilon \rightarrow 0$ we obtain the $\beta$-functions
\begin{align}
\label{eq:betas}
\beta(m^2) \equiv \frac{d m^2}{d\log\rg} &= \frac{\gsq m^2}{(4\pi)^2}, \quad \quad
\beta(\gsq) \equiv \frac{d \gsq}{d\log\rg} = \frac{3 \gsqsq }{(4\pi)^2}.
\end{align}

Shifting the field
$\phi\to\phi + v$
by the homogeneous background $v$, produces the tree-level potential in Eq.~\eqref{eq:Vtree}, and
the Lagrangian for the quantum field $\phi$ in the shifted theory reads
\begin{align}
\label{eq:shift1}
\mathcal{L} &=
    \frac{1}{2} (\partial_\mu \phi)^2
  + \frac{1}{2} \underbrace{\Big(
      m^2
    + \frac{1}{2} \gsq v^2 \Big)}_{\equiv M^2(v)} \phi^2
  + \underbrace{\Big(
     m^2 v
    + \frac{1}{6} \gsq v^3  \Big)}_{{\rm d}V_{\rmii{tree}}/{\rm d}v} \phi
    + \underbrace{\Big( \gsq v \Big)}_{V_{\phi^3}} \phi^3
  + \frac{1}{4!} \gsq \phi^4
  \;.
\end{align}
The coefficient of the linear term vanishes at the tree-level minimum.
The background field-dependent mass parameter $M^2(v)$ of the quantum field $\phi$ in
the shifted theory can be used to derive one-loop quantum corrections to the effective potential,
the Coleman-Weinberg potential~\cite{Coleman:1973jx}.
The effective potential is the generator of all $n$-point 1-particle-irreducible correlation functions $\langle \phi^n \rangle$ with zero external momentum,
\begin{align}
  V_{\rmii{CW}} = \sum_{n=1}^{\infty} \frac{\langle \phi^n \rangle}{n!} v^n
  \;.
\end{align}
To compute this, we
employ a trick~\cite{Lee:1974fj}: we shift $v\to v-\omega$ and
take a derivative with respect to $\omega$ at $\omega = v$.
This equals \textit{a single tadpole diagram} in the theory with background $v-\omega$ (quantities of this theory are denoted by tilde)
\begin{align}
\frac{{\rm d} V_{\rmii{CW}}}{{\rm d}\omega}\Big|_{\omega=s} =
  - \widetilde{\langle \phi \rangle} =
  \frac{1}{2} \Big( \frac{e^\gammaE \rg^2}{4\pi} \Big)^\epsilon
  \int \frac{{\rm d}p^D}{(2\pi)^D}
  \frac{\widetilde{V}_{v^3}}{p^2 + \widetilde{M}^2(v)}
\;,
\end{align}
where
$\widetilde{M}^2(v)$ and
$\widetilde{V}_{\phi^3}$ are those in Eq.~\eqref{eq:shift1}
with $v\to v-\omega$ replaced.
Integration over momentum is performed in dimensional regularisation in $D=4-2\epsilon$ dimensions in the Modified Minimal Subtraction (\MSbar) scheme~\cite{Peskin:1995ev}.
Since
$\widetilde{V}_{v^3} = \frac{{\rm d}}{{\rm d}\omega} \widetilde{M}^2(v)$,
an integration over omega retrieves (up to a constant)
\begin{align}
\label{eq:cw}
V_{\rmii{CW}} &=
  \frac{1}{2} \Big( \frac{e^\gammaE \rg^2}{4\pi} \Big)^\epsilon
  \int \frac{{\rm d}p^D}{(2\pi)^D} \ln \Big(p^2 + M^2(v) \Big)
  =
  -\frac{1}{2}
  \Big( \frac{\rg^2 e^\gammaE}{4\pi} \Big)^\epsilon
  \frac{ \Big(M^2(v) \Big)^\frac{D}{2}}{(4\pi)^{\frac{D}{2}}} \frac{\Gamma(-\frac{D}{2})}{\Gamma(1)} \nn \\ 
&= \frac{1}{(4\pi)^2} \frac{M^4(v)}{4} \bigg( -\frac{1}{\epsilon} - \frac{3}{2} + \ln \Big(\frac{M^2(v)}{\rg^2} \Big)  \bigg) + \mathcal{O}(\epsilon)
  \;,
\end{align}
where
$\Gamma$ is Euler's gamma function and
$\gammaE$ is the Euler-Mascheroni constant.
Integrals of this type are computed by employing standard parametrisation tricks \`a la Feynman and Schwinger~\cite{Peskin:1995ev}.
Divergent $1/\epsilon$ poles are cancelled by the tree-level counterterm contribution
\begin{align}
V_{\text{CT}} =\frac{1}{2} \delta m^2 v^2+ \frac{1}{4!} \delta \gsq v^4
\end{align}
and the remaining finite part yields $V_\rmi{CW}(v)$ of Eq.~\eqref{eq:VCWbar}.
An alternative, and perhaps easier derivation of the $\beta$-functions follows from the Renormalisation Group equation -- or Callan-Symanzik equation
-- that requires that the effective potential is scale independent $dV_{\text{eff}}/d\rg =0$.
By applying the chain rule, this leads to 
\begin{align} \label{eq:rge_appendix}
-\frac{\partial}{\partial \rg} V_\rmi{CW} = \Big( \beta(\gsq) \frac{\partial}{\partial \gsq} + \beta(m^2) \frac{\partial}{\partial m^2} - \gamma v \frac{\partial}{\partial v} \Big) V_{\text{tree}},
\end{align}
where we equate terms at $\mc{O}(\hbar)$. The anomalous dimension $\gamma$ 
is given by $\gamma = \frac{1}{2} (\Pi')^{-1} \rg \frac{\partial}{\partial \rg} \Pi'$,
where $\Pi' \equiv \frac{d}{dk^2} \Pi_2$, i.e.\ the part quadratic in external momentum $k$ of the self-energy function $\Pi_2$. 
At one-loop order the self-energy in this theory is momentum independent (the leading momentum dependence arises only at two-loop from the sunset topology diagram) and 
hence $\gamma$ vanishes. Then, requiring that the LHS and RHS are equal for any $v$ allows one to solve for the $\beta$-functions, resulting in Eqs.~\eqref{eq:betas}. 

\paragraph{Thermal corrections}

At high temperature, the one-loop contribution is given by (from now on we denote $M^2(v) \to M^2$, keeping background field dependence implicit in our notation)
\begin{align}
V_{\text{1-loop}} &= \frac{1}{2} \sumint{P} \ln(P^2 + M^2) = V_{\text{CW}} + V_T\left(\frac{M}{T}\right) ,
\end{align}
where $D$-dimensional integration at zero temperature has been replaced by a sum-integral, utilising the Matsubara or imaginary time formalism of thermal quantum field theory~\cite{Kapusta:2006pm,Laine:2016hma}: 
\begin{align}
\label{eq:Tint}
\Tint{P} &\equiv T \sum_{\omega_n} \int_p
\;, \quad\quad
\int_p \equiv \Big( \frac{\rg^{2}e^\gamma}{4\pi} \Big)^\epsilon \int \frac{{\rm d}^{d}p}{(2\pi)^d},
\end{align}
where $d=D-1$ and
Euclidean four-momentum is defined as
$P \equiv (\omega_n,\vec{p})$  
with the bosonic Matsubara frequency
$\omega_n = 2\pi n T$, where $n$ is integer.
In addition, we have divided\footnote{For a specific derivation, see \cite{Dolan:1973qd,Quiros:1999jp}, and for a more generic formula to separate vacuum and thermal contributions -- by replacing a thermal sum by a complex contour integral including the Bose distribution --  see Sec. 2.2 in Ref.~\cite{Laine:2016hma} or Sec. 3.4 in Ref.~\cite{Kapusta:2006pm}.}  the sum-integral into a temperature-independent Coleman-Weinberg piece and a thermal piece
\begin{align}
\label{eq:VT}
V_T(z) \equiv - T \int \frac{d^3 p}{(2\pi)^3} \ln \Big(1 + n_B(E_p,T) \Big) = T^4 \bigg( \frac{z^2}{24} - \frac{z^3}{12\pi} - \frac{z^4}{4(4\pi)^2}\ln\Big(\frac{z^2}{a_b} \Big) + \mathcal{O}(z^6) \bigg),
\end{align}
where $z\equiv M/T$, the Bose-distribution is $n_B(E_p,T) = 1/(e^{E_p/T}-1)$ with $E_p = \sqrt{p^2 + M^2}$ and $a_b \equiv (4\pi)^2 \text{Exp}(\frac{3}{2}-2\gamma)$.
In Eq.~\eqref{eq:VT} we have used the high-$T$ expansion  ($M/T \ll 1$) and dropped a constant term, independent of $z$.
Note that $V_T$ is UV-finite, and does not contain logarithms of renormalisation scale.

Merely evaluating Eq.~\eqref{eq:VT} is not sufficient to correctly capture the $\mc{O}(g^3)$ contribution to the thermal effective potential, as an infinite set of (daisy) diagrams contribute at this order. 
A minimal prescription for the necessary resummation~\cite{Arnold:1992rz} is to add the correction $V_T \rightarrow V_T + V_{\text{daisy}}$, where the daisy term -- associated to the cubic term that is non-analytic in mass squared -- reads
\begin{align}
\label{eq:daisy-resum}
V_{\text{daisy}} = -\frac{T}{12\pi} \bigg( (M^2 + \Pi_T)^{\frac{3}{2}} -(M^2)^{\frac{3}{2}} \bigg),
\end{align}
where $M^2 + \Pi_T$ is the \textit{resummed} background field-dependent mass parameter. 
This form is a result of the resummation of the mass of the zero Matsubara mode alone (by adding and subtracting the one-loop thermal mass correction to reorganise the perturbative expansion). Note that the second term in Eq.~\eqref{eq:daisy-resum} just removes the corresponding cubic term with unresummed mass in Eq.~\eqref{eq:VT}.     
In total, the one-loop thermal effective potential then reads
\begin{align}
\label{eq:Vloop1-usual}
V^{\text{1-loop}}_{\text{thermal}} = V_{\text{tree}} + V_{\text{CT}} + V_{\text{CW}} + V_T + V_{\text{daisy}}.
\end{align}
Note in particular, that in the high-$T$ expansion, the $\ln[M^2(v)]$ cancels between $ V_{\text{CW}} + V_T$, leaving only temperature and renormalisation scale remaining inside logarithms.
The thermal part $V_T$ can also be computed without a high-$T$ expansion, by numerically evaluating the integral in Eq.~\eqref{eq:VT}.
While the one-loop thermal part $V_T$ is explicitly free of logarithms of renormalisation scale, at leading order in the high-$T$ expansion it matches the tree-level part in power counting, and hence the running of parameters inside $V_T$ is sizeable.  

We find it helpful to write the thermal effective potential in an alternative form by separating the soft (zero mode) and hard (non-zero mode) contributions
\begin{align}
V_{\text{1-loop}} &= \frac{1}{2} \bigg( \underbrace{ T \int_p \ln(p^2+M^2+\Pi_T)}_{V_{\text{soft}}} +  \underbrace{\sumint{P}' \ln(P^2 + M^2)}_{V_{\text{hard}}} \bigg),
\end{align}
where 
\begin{align}
\Tint{P}' \equiv T \sum_{\omega_n \neq 0} \int_p.
\end{align}
The soft mode has a resummed mass and the soft piece is UV finite and reads simply
\begin{align}
V_{\text{soft}} &= - T \frac{(M^2 + \Pi_T)^{\frac{3}{2}}}{12 \pi}.
\end{align}
In the high-$T$ expansion, the (UV divergent) hard mode part reads
\begin{align}
\label{eq:V-hard}
V_{\text{hard}} &= \frac{M^2 T^2}{24} - \frac{M^4}{(4\pi)^2} \frac{1}{4} \Big( \frac{1}{\epsilon} + L_b(\mu) \Big) + \mathcal{O}(M^6/T^2)
\end{align}
and we can write (note that this form is equal to the more common form of Eq.~\eqref{eq:Vloop1-usual})
\begin{align}
\label{eq:Vloop1-soft-hard}
V^{\text{1-loop}}_{\text{thermal}} = V_{\text{tree}} + V_{\text{CT}} + V_{\text{soft}} + V_{\text{hard}},
\end{align}
and explicitly
\begin{align}
\label{eq:Vloop1-soft-hard-explicit}
V^{\text{1-loop}}_{\text{thermal}} = V_{\text{tree}}  + \frac{M^2 T^2}{24} - T \frac{(M^2 + \Pi_T)^{\frac{3}{2}}}{12 \pi} - \frac{M^4}{(4\pi)^2} \frac{L_b(\mu)}{4},
\end{align}
which completes the derivation of Eq.~\eqref{eq:Veff_hack}.
Note that $V_{\text{CT}}$ cancels the divergent part of $V_{\text{hard}}$.

The particular form of the thermal effective potential in Eq.~\eqref{eq:Vloop1-soft-hard} helps to illuminate the connection to \textit{dimensionally reduced} 3d EFT~\cite{Kajantie:1995dw, Braaten:1995cm}.
At one-loop order, this connection takes the form
\begin{align}
V^{\text{1-loop}}_{\text{thermal}} = \underbrace{V_{\text{tree}}  + \frac{M^2 T^2}{24} + \frac{M^4}{(4\pi)^2} \frac{L_b(\mu)}{4}}_{\sim T \; V^{\text{tree}}_{\text{3d}}} - \underbrace{T \frac{(M^2 + \Pi_T)^{\frac{3}{2}}}{12 \pi}}_{\sim T \; V^{\text{1-loop}}_{\text{3d}}}.
\end{align}
The full $\mc{O}(g^4)$ effective potential for this theory, given in Eq.~\eqref{eq:veff-NNLO}, can be constructed using high-temperature dimensional reduction~\cite{Farakos:1994kx,Braaten:1995cm,Andersen:1997zx,Gould:2021dzl,Schicho:2021gca}.
For this, one needs the two-loop effective potential within the 3d EFT,
\begin{align}
\label{eq:3d-veff}
V_{\text{thermal}}=T \; V^{\text{3d}}_{\text{eff}} &= T \; \bigg\{
\frac{1}{2} m^2_{3}(\mu_3) v_{3}^{2}
  + \frac{1}{4!} g^2_{3} v_{3}^{4}
  - \frac{1}{3(4\pi)} (M^2_{3})^{\frac{3}{2}}\nonumber \\
  &+ \frac{1}{(4\pi)^2} \bigg(
     \frac{1}{8} g^2_{3} M^2_{3}
    - \frac{1}{24} g^4_{3} v^2_{3}
      \Big[1+ 2 \ln \Big( \frac{\mu_3}{3 M_{3}} \Big) \Big]
  \bigg)
  \bigg\} \;,
\end{align}
where for convenience we have defined
\begin{align}
M^2_{3} &= m^2_3(\mu_3) + \frac{1}{2}g^2_3 v^2_3.
\end{align}
The 3d effective parameters are needed up to the same $\mc{O}(g^4)$ order:
\begin{align} \label{eq:g3sq}
g^2_3 &= T \Big( g^2(\mu) - \frac{3}{2(4\pi)^2} g^4 L_b(\mu) \Big), \\
\label{eq:m3sq}
m^2_3(\mu_3) &= m^2(\mu) + \frac{1}{24} g^2(\mu) T^2 \nonumber \\
&\quad
- \frac{1}{(4\pi)^2} \left(
\frac{1}{2} g^2 m^2 L_b(\mu)
+ \frac{1}{16}  g^4 T^2 L_b(\mu) 
+ \frac{1}{6} g^4_3 \Big[ c + \ln \Big( \frac{3 T}{\mu_3} \Big) \Big] \right), \\
v_3 &= \frac{v}{\sqrt{T}}. \label{eq:v3}
\end{align}
Expanding Eq.~\eqref{eq:3d-veff} up to $\mc{O}(g^4)$ results in Eq.~\eqref{eq:veff-NNLO}, while the unexpanded form contains a subset of higher-order resummations.

Note that the renormalisation scale $\mu$ cancels at $\mathcal{O}(g^4)$ in Eqs.~\eqref{eq:g3sq} to \eqref{eq:v3}: the running of the leading terms is cancelled by the explicit $\mu$-dependence of the $L_b(\mu)$ terms.
The residual $\mu$-dependence is $\mathcal{O}(g^6)$.
In addition to these UV logarithms, there arise IR logarithms within the full theory.
These match against corresponding logarithms determining the running of parameters within the 3d EFT, allowing one to replace $\mu\to\mu_3$ within all IR logarithms~\cite{Farakos:1994kx}; see Eq.~\eqref{eq:m3sq}.
This is characteristic of EFTs more generally~\cite{Manohar:2018aog}.
Finally, note that within Eq.~\eqref{eq:3d-veff} for the effective potential the $\mu_3$-dependence of the tree-level mass parameter cancels against that of the explicit two-loop term, leaving a residual $\mu_3$-dependence at $\mathcal{O}(g^5)$.

The advantages of the EFT approach are even clearer at the next order~\cite{Rajantie:1996np}.
For a complete determination of the $\mathcal{O}(g^5)$ effective potential in this theory, one needs to compute three-loop diagrams.
However, only ordinary $d$-dimensional integrals within the EFT are needed at this order, and no new sum-integrals, i.e.\ Eq.~\eqref{eq:3d-veff} needs extending but not Eqs.~\eqref{eq:g3sq} to \eqref{eq:v3}.


%
\bibliographystyle{JHEP}
\bibliography{refs}

\end{document}